\documentclass[%
 aip,
 amsmath,amssymb,
 reprint,%
 citeautoscript,
]{article}

\usepackage[final, nonatbib]{ml4mol}

\usepackage{graphicx}%
\usepackage{dcolumn}%
\usepackage{bm}%
\usepackage{braket}
\usepackage{xcolor}
\usepackage{appendix}
\usepackage{tabularx}
\usepackage[
    natbib=true,
    style=numeric,
    sorting=none
    ]{biblatex}
\usepackage{float}

\usepackage{amsmath}
\usepackage{mathtools} 
\usepackage{braket}
\usepackage{xcolor}
\usepackage{subfigure}

\usepackage{wrapfig}

\usepackage[capitalize]{cleveref}
\usepackage[mathscr]{eucal}
\usepackage[utf8]{inputenc}
\usepackage[T1]{fontenc}
\usepackage{siunitx}
\usepackage{multirow}

\DeclarePairedDelimiter\paran{ ( }{ ) }
\newcommand\footnoteref[1]{\protected@xdef\@thefnmark{\ref{#1}}\@footnotemark}

\begin{document}

\title{
 Multi-task learning for  electronic structure to predict and explore molecular potential energy surfaces
} %

\author{
Zhuoran Qiao \\
California Institute of Technology\\
Pasadena, CA 91125 \\
\texttt{zqiao@caltech.edu}
\And
Feizhi Ding\\
Entos, Inc.\\ 
Los Angeles, CA 90027 \\
\texttt{feizhi@entos.ai}
\And
Matthew Welborn \\
Entos, Inc.\\ 
Los Angeles, CA 90027 \\
\texttt{matt@entos.ai}
\And
Peter J. Bygrave\\
Entos, Inc.\\ 
Los Angeles, CA 90027 \\
\texttt{peter@entos.ai}
\And
Daniel G. A. Smith\\
Entos, Inc.\\ 
Los Angeles, CA 90027 \\
\texttt{daniel@entos.ai}
\And
Animashree Anandkumar \\
California Institute of Technology\\
Pasadena, CA 91125 \\
NVIDIA\\
Santa Clara, CA 95051\\
\texttt{anima@caltech.edu}
\And
Frederick R. Manby \\
Entos, Inc.\\ 
Los Angeles, CA 90027 \\
\texttt{fred@entos.ai}
\And
Thomas F. Miller III\\
California Institute of Technology\\
Pasadena, CA 91125 \\
Entos, Inc.\\ 
Los Angeles, CA 90027 \\
\texttt{tfm@caltech.edu, tom@entos.ai} \\
}

\date{\today}%
\maketitle

\begin{abstract}
We refine the OrbNet model to accurately predict energy, forces, and other response properties for  molecules using a graph neural-network architecture based on features from low-cost approximated quantum operators in the symmetry-adapted atomic orbital basis. The model is end-to-end differentiable due to the derivation of analytic gradients for all electronic structure terms, and is shown to be transferable across chemical space due to the use of domain-specific features. The learning efficiency is improved by incorporating physically motivated constraints on the electronic structure through multi-task learning. The model outperforms existing methods on energy prediction tasks for the QM9 dataset and for molecular geometry optimizations on conformer datasets, at a computational cost that is thousand-fold or more reduced compared to conventional quantum-chemistry calculations (such as density functional theory) that offer similar accuracy.
\end{abstract}

\begin{refsection}

\section{Introduction}

Quantum chemistry calculations -  most commonly those obtained using density functional theory (DFT) - provide a level of accuracy that is important for many chemical applications but at a computational cost that is often prohibitive.  As a result, machine-learning efforts have focused on the prediction of molecular potential energy surfaces, using both physically motivated  features \cite{Bartok2010,ramakrishnan2015big,Nguyen2018,Smith2017,smiti_transfer_2018,mobml2} and neural-network-based representation learning \cite{gat, yang2019analyzing, schutt2017schnet,gilmer2017neural, unke2019physnet, cormorant, DimeNet}. 
Despite the success of such methods in predicting energies on various benchmarks, the generalizability of deep neural network models across chemical space and for out-of-equilibrium geometries is less investigated.

In this work, we demonstrate an approach using features from a low-cost electronic-structure calculation in the basis of symmetry-adapted atomic orbitals (SAAOs) with a deep neural network architecture (OrbNet).  The model has previously been shown to predict the molecular energies with DFT accuracy for both chemical and conformational degrees of freedom, even when applied to systems significantly larger than the training molecules \cite{orbnet1}. 
To improve learning efficiency, we introduce a multi-task learning strategy in which  OrbNet  is trained with respect to both molecular energies and other computed properties of the quantum mechanical wavefunction.  Furthermore, we introduce and numerically demonstrate the analytical gradient theory for OrbNet, which is essential for the calculation of inter-atomic forces and other response properties, such as dipoles and linear-response excited states.  

\section{Method}

\subsection{OrbNet: Neural message passing on SAAOs with atomic and global attention}

\begin{figure*}
    \centering
    \includegraphics[width=0.6\textwidth]{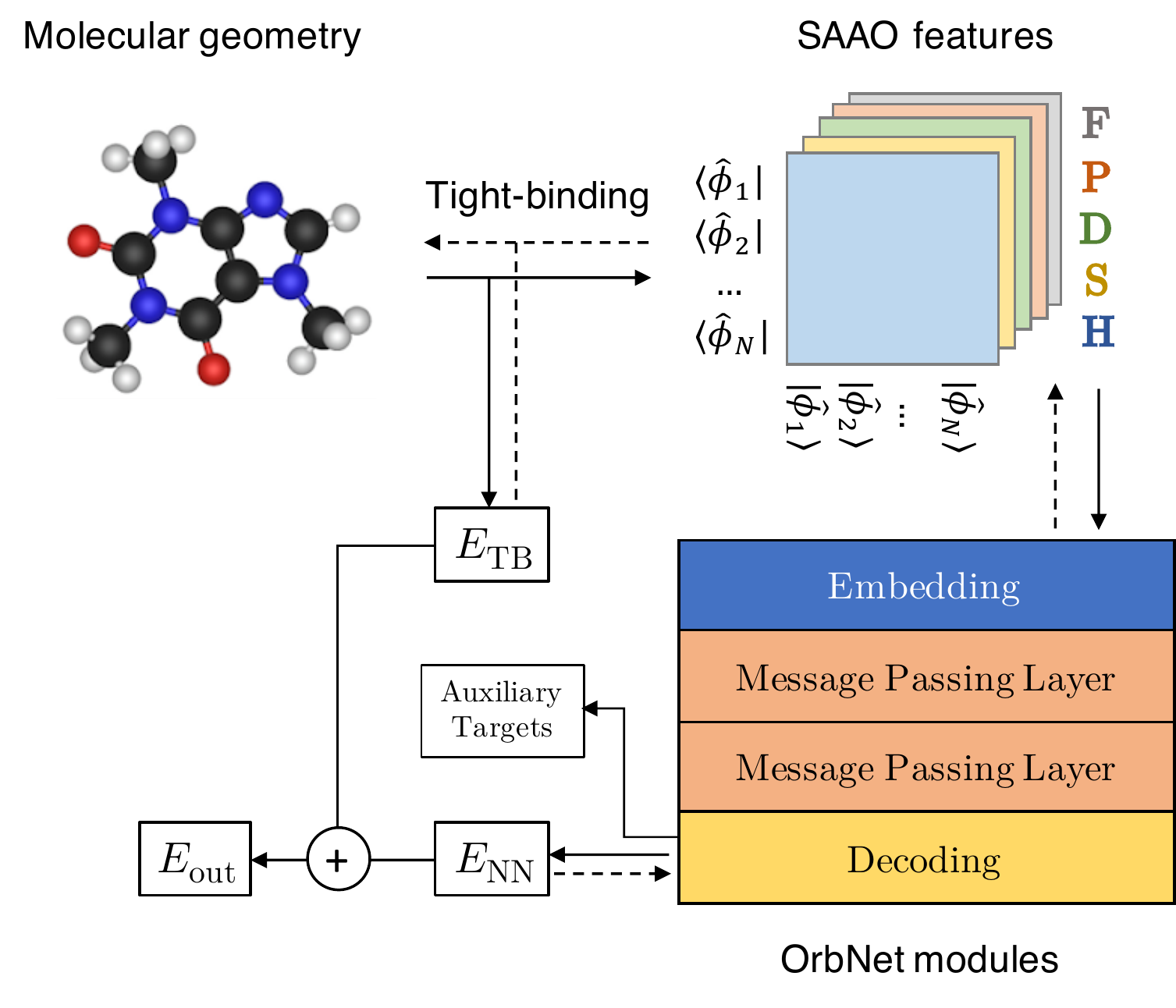}
    \caption{
  Schematic of the employed OrbNet deep-learning approach. 
  A low-cost extended tight-binding calculation is performed on the molecular system, generating the approximate energy $E_{\textrm{TB}}$ and the SAAO feature matrices. 
  The SAAO feature matrices are passed to the OrbNet deep neural network modules, yielding the machine-learned correction to the total energy, $E_{\textrm{NN}}$, as well as auxiliary atom-specific 
  target properties. 
  Dashed arrows indicate components that must be analytically differentiated for the gradient calculation. 
Additional details are provided in 
Appendix \ref{sec:model}. 
  }
    \label{fig:gnn}
\end{figure*}

In this work, the molecular system is encoded as graph-structured data with features obtained from a low-cost tight-binding calculation, following  Qiao et al \cite{orbnet1}.  We employ features obtained from matrix elements of approximated quantum operators of an extended tight-binding method (GFN-xTB\cite{gfn1}), evaluated in the symmetry-adapted atomic orbital (SAAO) basis. 
Specifically, the Fock ($\mathbf{F}$), density ($\mathbf{P}$), orbital centroid distances ($\mathbf{D}$), core Hamiltonian ($\mathbf{H}$), and overlap ($\mathbf{S}$) matrices are used as the input features, 
with node features corresponding to diagonal SAAO matrix elements $X_u=[F_{uu}, P_{uu}, H_{uu}]$ and edge features corresponding to off-diagonal SAAO matrix elements $X^\mathrm{e}_{uv}=[F_{uv}, D_{uv}, P_{uv}, S_{uv}, H_{uv}]$. Fig.~\ref{fig:gnn} summarizes the deep-learning approach, and additional details are provided in Appendix \ref{sec:model}.

The feature embedding and neural message-passing mechanism employed for the node and edge attributes is largely unchanged from Ref.~\cite{orbnet1}.  
However, to enable multi-task learning and to improve the learning capacity of the model, 
we introduce atom-specific attributes, $\mathbf{f}_A^{l}$, and  global molecule-level attributes, $\mathbf{q}^l$, where $l$ 
is the message passing layer index and $A$ is the atom index.
The whole-molecule and atom-specific attributes  allow for the prediction of auxiliary targets (Fig.~\ref{fig:gnn}) through multi-task learning, thereby providing physically motivated constraints on the electronic structure of the molecule that can be used to  refine the representation at the SAAO level.

For the prediction of both the electronic energies and the auxiliary targets, only the final atom-specific attributes, $\mathbf{f}_A^L$, are employed, since they self-consistently incorporate the effect of the whole-molecule and node- and edge-specific attributes.
The electronic energy is obtained by combining the approximate energy $E_{\mathrm{TB}}$ from the extended tight-binding calculation and the model output $E_{\mathrm{NN}}$, the latter of which is a one-body sum over atomic contributions; 
the atom-specific auxiliary targets $\mathbf{d}_A$ are predicted from the same attributes.
\begin{align} 
    \hat{E}_{\textrm{out}} &= E_\mathrm{TB} + E_\mathrm{NN} = E_\mathrm{TB} + \sum_{A} \epsilon_A = E_\mathrm{TB} + \sum_{A} [\mathrm{Dec}(\mathbf{f}_A^L) + E^\text{c}_{A}] \\
    \hat{\mathbf{d}}_A &= \mathrm{Dec}^{\textrm{aux}}(\mathbf{f}_A^L)
\end{align}

Here, the energy decoder  $\mathrm{Dec}$ and the auxiliary-target decoder  $\mathrm{Dec}^{\textrm{aux}}$ are residual neural networks \cite{resnet} built with fully connected and normalization layers, and $E^\text{c}_{A}$ are element-specific, constant shift parameters for the isolated-atom contributions to the total energy.

The GradNorm algorithm \cite{chen2018gradnorm} is used to adaptively adjust the weight of the auxiliary target loss based on the gradients of the last fully-connected layer before the decoding networks.

\subsection{End-to-end differentiability: Analytic gradients}

The model is constructed to be end-to-end differentiable by employing input features (i.e., the SAAO matrix elements) that are smooth functions of both atomic coordinates and external fields. 
We derive the analytic gradients of the total energy $E^{\textrm{out}}$ with respect to the atom coordinates, and we employ local energy minimization with respect to molecular structure as an exemplary task to demonstrate the quality of the learned potential energy surface (Section \ref{sec:geo_opt}).

Using a Lagrangian formalism \cite{lee2019analytical,schutz2004analytical}, the analytic gradient of the predicted energy with respect to an atom coordinate $x$ can be expressed in terms of contributions from the tight-binding model, the neural network, and additional constraint terms:
\begin{equation}
\frac{d E_{\text{out}}}{d x} = \frac{d E_{\text{TB}}}{d x} + \sum_{\mathbf{f} \in \Set{\mathbf{F}, \mathbf{D}, \mathbf{P}, \mathbf{S}, \mathbf{H}}} \text{Tr}\left[ \frac{\partial E_{\text{NN}}}{\partial \mathbf f} \frac{\partial \mathbf f}{\partial x} \right]  + \text{Tr} [\mathbf W \frac{\partial \mathbf{S}^{\mathrm{AO}}}{\partial x}]   +  \text{Tr} [\mathbf  z \frac{\partial \mathbf{F}^{\mathrm{AO}}}{\partial x}].  \label{eq:L_grad}
\end{equation}
Here, the third  and fourth terms on the right-hand side are gradient contributions from the orbital orthogonality constraint and the Brillouin condition, respectively, where $\mathbf{F}^{\mathrm{AO}}$ and $\mathbf{S}^{\mathrm{AO}}$ are the Fock matrix and orbital overlap matrix in the atomic orbital (AO) basis.
Detailed expressions for $\frac{\partial \mathbf f}{\partial x}$, $\mathbf{W}$, and $\mathbf{z}$  are provided in Appendix \ref{sec:grad}. The  tight-binding gradient $\frac{d E_{\text{TB}}}{d x}$ for the  GFN-xTB model has been previously reported \cite{gfn1}, and the neural network gradients with respect to the input features $\frac{\partial E_{\text{NN}}}{\partial \mathbf f}$ are obtained using reverse-mode automatic differentiation \cite{pytorch}. 

\subsection{Auxiliary targets from density matrix projection}

The utility of graph- and atom-level auxiliary tasks to improve the generalizability of the learned representations for molecules has been highlighted for learning molecular properties in the context of graph pre-training \cite{hu2019strategies, goh2018using} and multi-task learning \cite{DeepMoleNet}. Here, we employ 
multi-task learning with respect to the total molecular energy and atom-specific 
auxiliary targets.  
The atom-specific targets that we employ are similar to the features introduced in the DeePHF model \cite{deephf}, obtained by projecting the density matrix into a basis set that does not depend upon the identity of the atomic element, 
\begin{equation}
    \mathbf{d}^{A}_{nl} = [ \text{EigenVals}_{m,m'}([~{}^{\mathrm{O}}\!\mathscr{D}^A_{nl}]_{m,m'})|| \text{EigenVals}_{m,m'}([~{}^{\mathrm{V}}\!\mathscr{D}^A_{nl}]_{m,m'})]. 
\end{equation}
Here, the projected density matrix is given by $[~{}^{\mathrm{O}}\mathscr{D}^A_{nl}]_{m,m'}= \sum_{i\in\textrm{occ}} \langle \alpha^{A}_{nlm}|\psi_i \rangle \langle \psi_i | \alpha^{A}_{nlm'}\rangle$, and the  projected valence-occupied density matrix is given by  $[~{}^{\mathrm{V}}\mathscr{D}^A_{nl}]_{m,m'}= \sum_{j\in\textrm{valocc}} \langle \alpha^{A}_{nlm}|\psi_j \rangle \langle \psi_j | \alpha^{A}_{nlm'}\rangle$, where $|\psi_{\{i,j\}} \rangle$ are molecular orbitals from the reference DFT calculation, $|\alpha^{A}_{nlm}\rangle$ is a basis function centered at atom $A$ with radial index $n$ and spherical-harmonic degree $l$ and order $m$.  The indices $i$ and $j$ runs over all occupied orbitals and valence-occupied orbital indices, respectively, and $||$ denotes a vector concatenation operation. The auxiliary target vector $\mathbf{d}_A$ for each atom $A$ in the molecule is obtained by concatenating $\mathbf{d}^{A}_{nl}$ for all $n$ and $l$. The parameters for the projection basis $|\alpha^{A}_{nlm}\rangle$ are described in Appendix \ref{sec:aux_proj}. 
Additional attributes, such as such as partial charges and reactivities, could also be naturally included within this framework.

\section{Results}
\label{sec:results}

We present results for molecular energy prediction and  geometry optimization tasks. All models are produced using the same set of hyperparameters and the training procedure  in Appendix \ref{sec:hyperparams}. 

\subsection{QM9 formation energy}
\label{sec:qm9}

We begin with a standard benchmark test of predicting molecular energies for the QM9 dataset, which consists of 133,885 organic molecules with up to nine heavy atoms at locally optimized geometries.
Table \ref{table:qm9} presents results from current work, as well as previously published results using   SchNet \cite{schutt2017schnet}, PhysNet \cite{unke2019physnet}, DimeNet \cite{DimeNet},  DeepMoleNet \cite{DeepMoleNet}, and OrbNet \cite{orbnet1}. 
The approach proposed in this work significantly outperforms existing methods in terms of both data efficiency and prediction accuracy in this dataset.  In particular, it is seen that the use of multi-task learning in the current study leads to significant improvement over the previously published OrbNet results, which already exhibited the smallest errors among published methods.

\begin{table*}[h]
\setlength{\extrarowheight}{0.1cm}
\small
\centering
\begin{tabularx}{\textwidth}{p{0.12\textwidth}p{0.09\textwidth}p{0.11\textwidth}p{0.11\textwidth}p{0.15\textwidth}p{0.1\textwidth}p{0.1\textwidth}}
\hline
Training size & SchNet & PhysNet & DimeNet &DeepMoleNet & OrbNet  & This work \\ \hline
25,000 & - & -& -& -& 11.6 & \textbf{8.08} \\ \hline
50,000 & 15 & 13 & - & - & 8.22 & \textbf{5.89}\\ \hline
110,000 & 14 & 8.2 & 8.02 & 6.1 & 5.01 & \textbf{3.87} \\ \hline
\end{tabularx}
\caption{MAEs  (reported in meV) for predicting the QM9 dataset of %
total molecular energies. The employed labels are the published values \cite{qm9} calculated at the B3LYP/6-31G(2df,p) level of theory. }
\label{table:qm9}
\end{table*}

\subsection{Molecular geometry optimizations}
\label{sec:geo_opt}


\begin{figure*}
    \centering
    \includegraphics[width=0.8\textwidth]{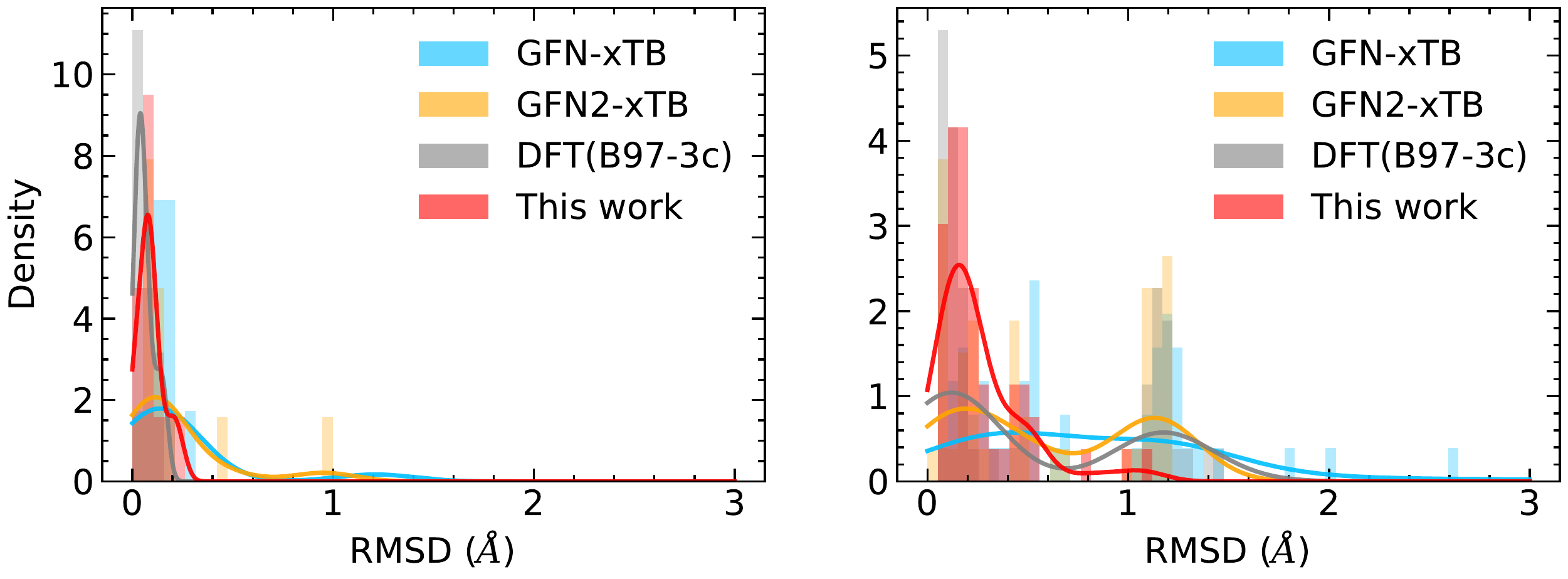}
    \setlength{\extrarowheight}{0.1cm}
    \small
    \begin{tabularx}{0.8\textwidth}{|c|c|c|c|c|X|}
    \hline
    \multirow{2}*{Method} & \multicolumn{2}{c}{Mean RMSD (\AA)} & \multicolumn{2}{|c|}{Incorrect geometries} & Time/step \\ \cline{2-6}
     & ROT34 & MCONF & ROT34 & MCONF & MCONF\\  \hline
    GFN-xTB  & 0.23 & 0.90 & 8\% & 52\% & < 1 s \\ \hline
    GFN2-xTB  & 0.21 & 0.60 & 8\% & 44\% & < 1 s  \\ \hline
    DFT (B97-3c) & \textbf{0.06} & 0.51 & \textbf{0\%} & 37\% & > 100 s \\ \hline
    This work & 0.09 & \textbf{0.26} & \textbf{0\%} & \textbf{6\%} & < 1 s  \\ \hline
    Ref.~DFT ($\omega$B97X-D3) & - & - & - & - & > 1,000 s  \\ \hline
    \end{tabularx}
    \caption{The molecular geometry optimization accuracy for the ROT34 (left) and MCONF (right) datasets, reported as the best-alignment root-mean-square-deviation (RMSD) 
    compared to the 
    reference DFT geometries at the $\omega$B97X-D3/Def2-TZVP level. The distribution of errors are plotted as histograms (with overlaying 
    kernel density estimations), and the table reports the mean errors and the percentage of optimized structures that correspond to  incorrect geometries (i.e., RMSD $>0.6$ Angstrom). 
    Timings correspond to the average cost for a single force evaluation for the MCONF dataset on a single Intel Xeon Gold 6130 @ 2.10GHz CPU core. 
    }
    \label{fig:geom}
\end{figure*}

A practical application of energy gradient (i.e., force) calculations is to  optimize molecule structures by locally minimizing the energy.  Here, we use this application as a test of the accuracy of the OrbNet potential energy surface in comparison to other widely used methods of comparable and greater computational cost.
Test are performed for the ROT34 \cite{rot34} and MCONF \cite{mconf} datasets, with initial structures that are locally optimized at the high-quality level of $\omega$B97X-D3/Def2-TZVP DFT with tight convergence parameters.
ROT34 includes conformers of 12 small organic molecules with up to 13 heavy atoms; MCONF includes 52 conformers of the melatonin molecule which has 17 heavy atoms. 
From these initial structures, we performed a local geometry optimization 
using the various energy methods, including OrbNet from the current work, the GFN semi-empirical methods \cite{gfn1,gfn2},  
and the relatively low-cost DFT functional B97-3c \cite{brandenburg2018b97}. The error in the resulting structure with respect to the reference structures optimized at the $\omega$B97X-D3/Def2-TZVP level was  computed as root mean squared distance (RMSD) following optimal molecular alignment.  This test investigates whether 
the potential energy landscape for each method is locally consistent with a high-quality DFT description.

Fig.~\ref{fig:geom} presents the resulting distribution of errors for the various methods over each dataset, with results summarized in the accompanying table.
It is clear that while the GFN semi-empirical methods provide a computational cost that is comparable to OrbNet, the resulting geometry optimizations are substantially less accurate, with a significant (and in some cases very large) fraction of the local geometry optimizations relaxing into structures that are inconsistent with the optimized reference DFT 
structures (i.e., with RMSD in excess of $0.6$ Angstrom).  In comparison to DFT using the B97-3c functional, OrbNet provides optimized structures that are of comparable accuracy for  ROT34 and that are significantly more accurate for MCONF; this should be viewed in light of the fact that OrbNet is over 100-fold less computationally costly.  On the whole, OrbNet is the best approximation to the reference DFT results,  at a computational cost that is over 1,000-fold reduced.

\section{Conclusions}

We extend the OrbNet deep-learning model through the use of multi-task learning and the development of the analytical gradient theory for calculating molecular forces and other response properties.  It is shown that multi-task learning leads to improved data efficiency, with OrbNet providing lower errors than previously reported deep-learning methods for the QM9 formation energy prediction task.  Moreover, it is shown that geometry optimizations on the OrbNet potentially energy surface provide accuracy that is significantly greater than that available from semi-empirical methods and that even outperform fully quantum mechanical DFT descriptions that are vastly more computationally costly.  The method is immediately applicable to other down-stream tasks.

\section*{Acknowledgement}
Z.Q. acknowledges the graduate research funding from Caltech. T.F.M. and A.A. acknowledge partial support from the Caltech DeLogi fund, and A.A. acknowledges support from a Caltech Bren professorship. The authors gratefully acknowledge NVIDIA, including Abe Stern and Tom Gibbs, for helpful discussions regarding GPU implementations of graph neural networks.

\printbibliography

\end{refsection}

\begin{refsection}

\newpage

\appendixpage

\appendix

\renewcommand{\thesection}{\Alph{section}}

\section{Dataset and computational details}
\label{sec:comp}

For results reported in Section \ref{sec:qm9}, we employ the QM9 dataset\cite{qm9} with pre-computed DFT labels. From this dataset, 3054 molecules were excluded 
as recommended in Ref.~\cite{qm9};
we sample 110000 molecules for training and 10831 molecules for testing. The training sets of 25000 and 50000 molecules are subsampled from the 110000-molecule dataset.

To train the model reported in Section \ref{sec:geo_opt}, we employ the published geometries 
from Ref.~\cite{orbnet1}, which include optimized and thermalized geometries of molecules up to 30 heavy atoms from the QM7b-T, QM9, GDB13-T, and DrugBank-T datasets.  We perform model training using  the dataset splits of Model 3 in Ref. \cite{orbnet1}. 
DFT labels are computed using the $\omega$B97X-D3 functional \cite{wb97xd3} with a Def2-TZVP AO basis set\cite{def2tzvp} and using density fitting\cite{Polly2004} for both the 
Coulomb and exchange integrals using the Def2-Universal-JKFIT basis set.\cite{def2-universal-jkfit} %

For results reported in Section \ref{sec:geo_opt}, we perform geometry optimization for the DFT, OrbNet, and GFN-xTB calculations by minimizing the potential energy using the BFGS algorithm with the Translation-rotation coordinates (TRIC) of Wang and Song\cite{tric}; geometry optimizations for GFN2-xTB are performed using the default algorithm in the \textsc{xtb} package \cite{xtbcode}.  All local geometry optimizations  are initialized from pre-optimized structures at the $\omega$B97X-D3/Def2-TZVP level of theory. For the B97-3c method, the mTZVP basis set\cite{brandenburg2018b97} is employed.

All DFT and GFN-xTB calculations are performed using  \textsc{Entos Qcore} \cite{entos}; GFN2-xTB calculation are performed using \textsc{xtb} package \cite{xtbcode}.

\begin{figure*}[h]
\centering
    \subfigure[GFN2-xTB v.s. reference, RMSD=1.2\AA]{
    \includegraphics[width=0.48\textwidth]{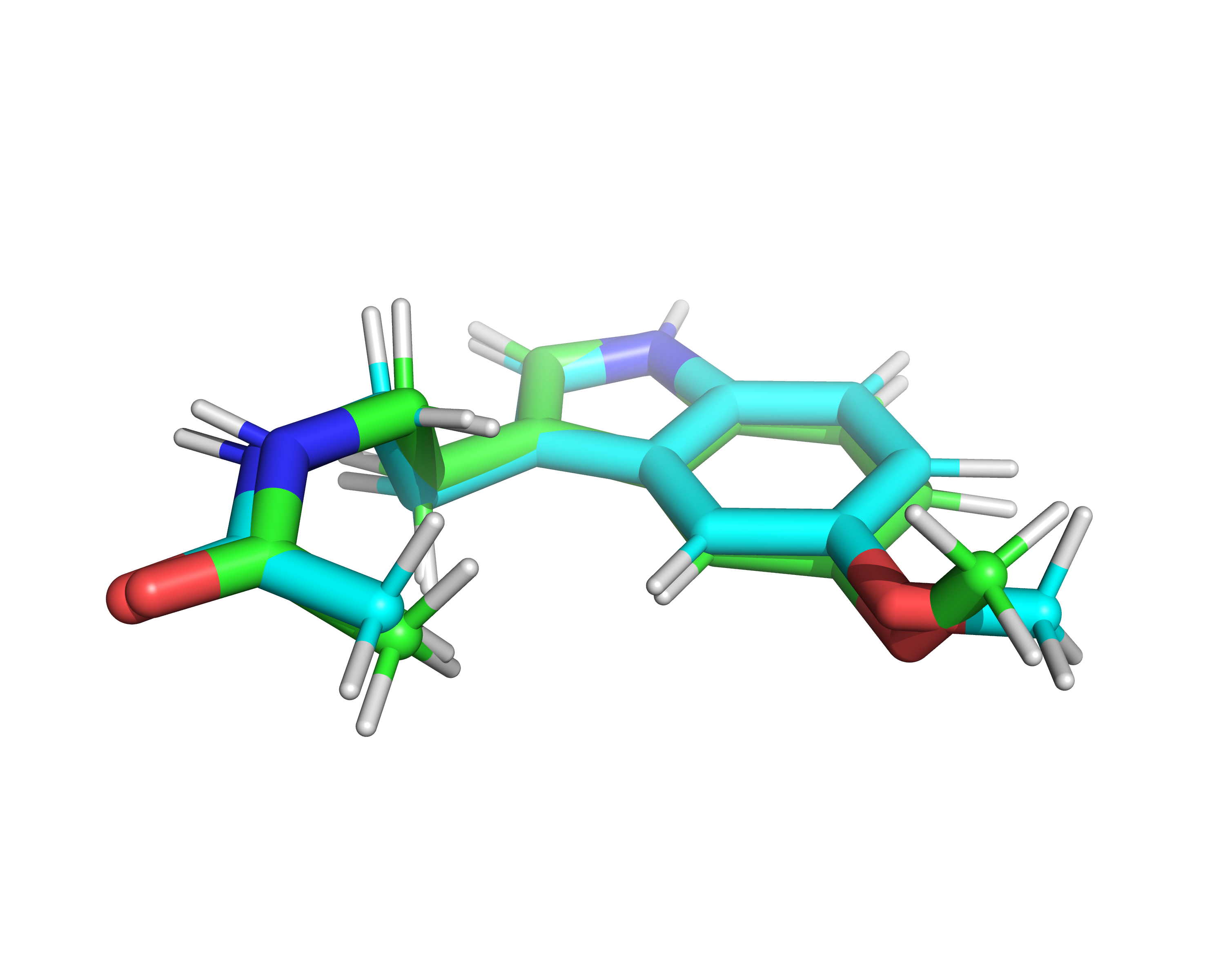}}
    \subfigure[OrbNet v.s. reference, RMSD=0.14\AA] {   
    \includegraphics[width=0.48\textwidth]{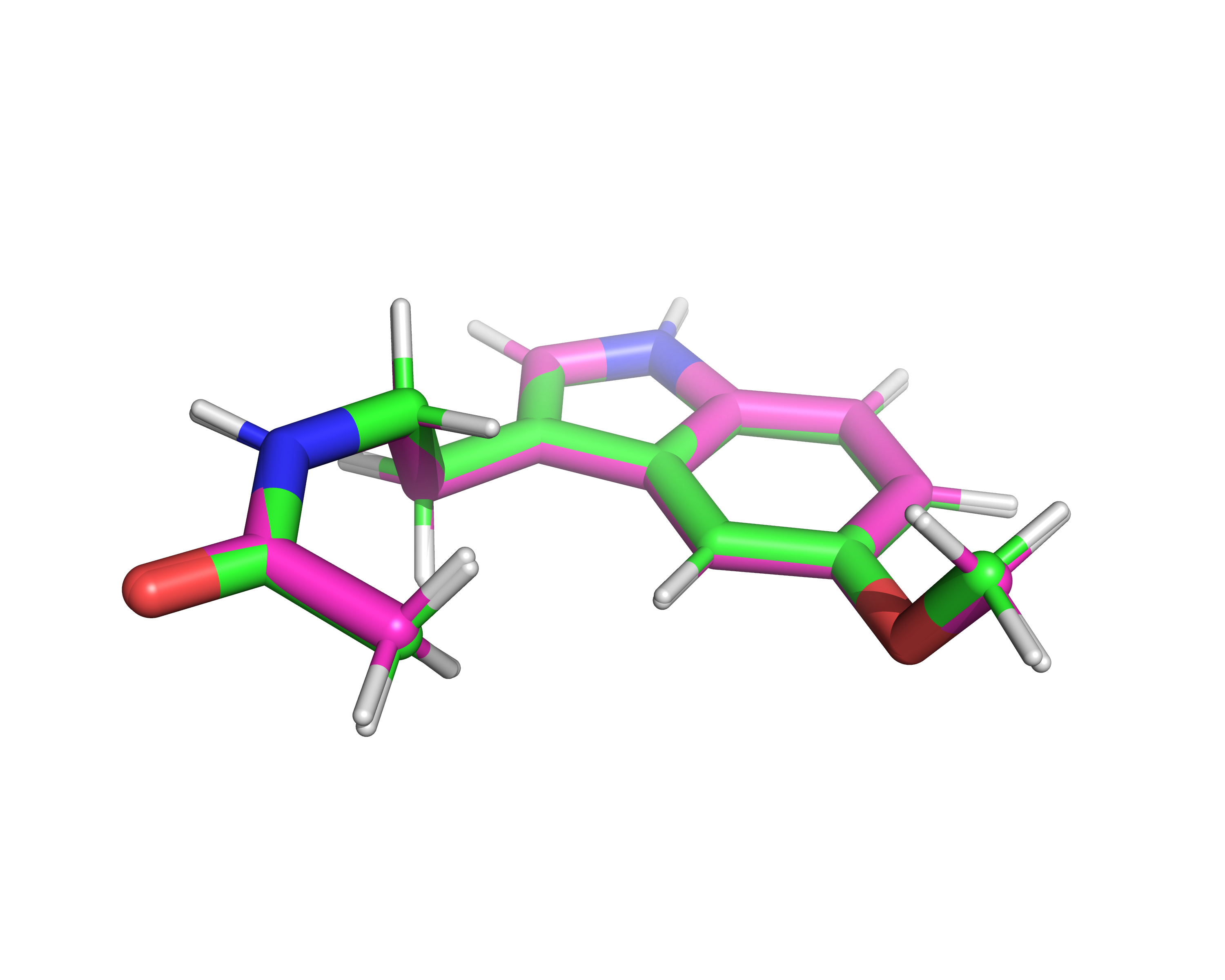}}
    \caption{Comparison of optimized structure of the conformer MCONF/41 from GFN2-xTB (cyan), OrbNet (magenta) and the reference DFT structure (green).}\label{fig:vis}
\end{figure*}

%



\section{Specification of OrbNet embedding, message passing \& pooling, and decoding layers}

\begin{figure*}
    \centering
    \includegraphics[width=\textwidth]{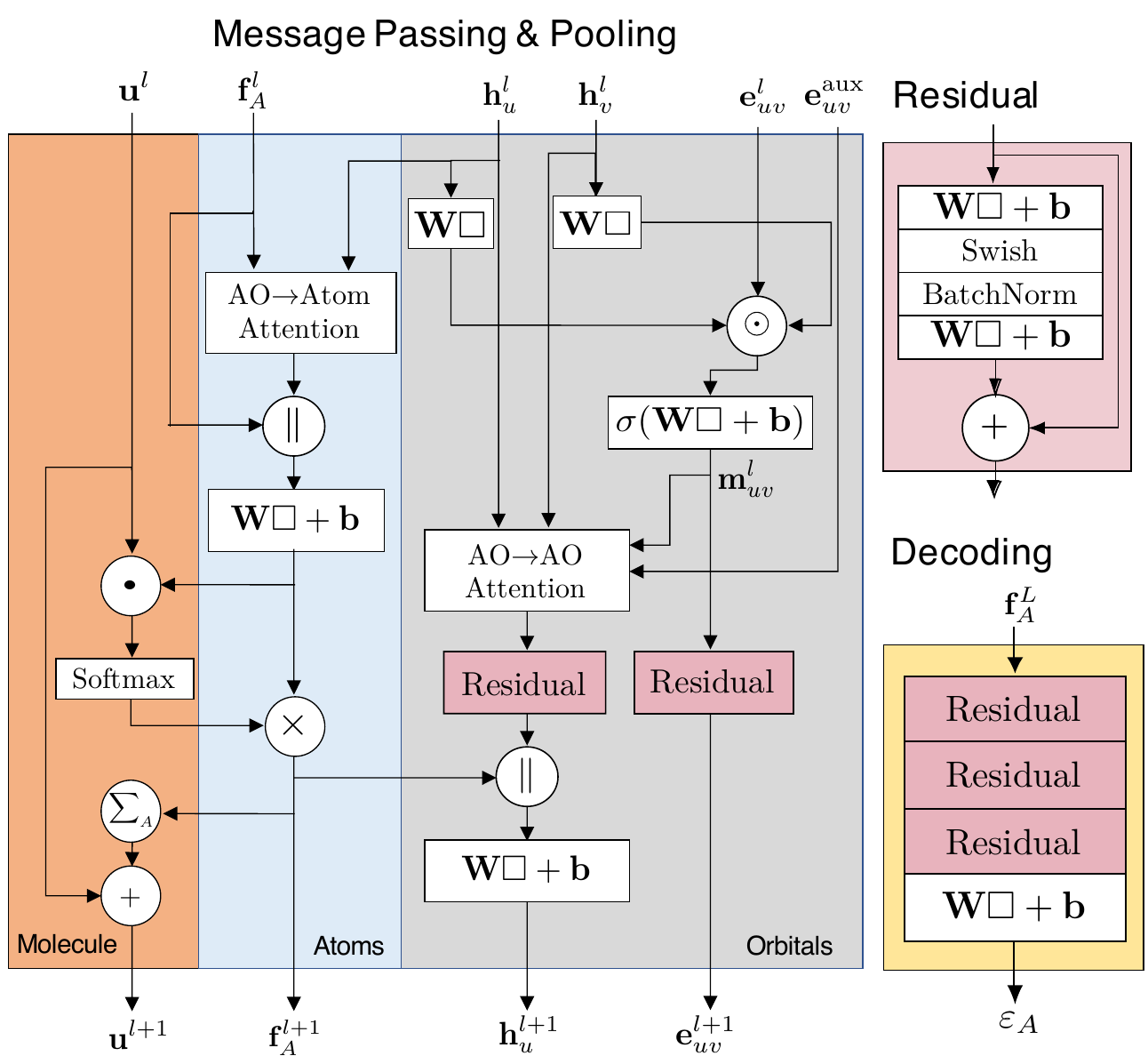}
    \caption{
Detail of a single message-passing and pooling layer (``Message Passing Layer'' in Fig. \ref{fig:gnn}), and a decoding network (``Decoding'' in Fig. \ref{fig:gnn}). At message passing and pooling layer $l+1$, the whole-molecule, atom-specific, node-specific, and edge-specific attributes are updated. The atom-specific attributes $\mathbf{f}_A^{l}$ are updated with input from 
node- and edge-specific attributes $\mathbf{h}_u^{l}$ and $\mathbf{e}_{uv}^{l}$ and likewise includes the back-propagation from the whole-molecule attributes; finally, the whole-molecule attributes $\mathbf{q}^{l}$ are updated with input from the atom-specific attributes. The final atom-specific attributes are passed into separate decoding networks to generate the energy prediction and auxiliary target predictions. A decoding network is composed of multiple residual blocks (``Residual'') and a linear output layer, as illustrated above.}
    \label{fig:mpl}
\end{figure*}

\label{sec:model}

We employ the feature embedding scheme introduced in OrbNet~\cite{orbnet1} where the SAAO feature matrices are transformed by radial basis functions, 
\begin{equation}
    \label{nrbf}
    \mathbf{h}^{\textrm{RBF}}_{u} = [\phi^\mathrm{h}_1(\tilde{X}_{u}),  \phi^\mathrm{h}_2(\tilde{X}_{u}), ..., \phi^\mathrm{h}_{n_\mathrm{r}}(\tilde{X}_{u})] 
\end{equation}
\begin{equation}
    \label{erbf}
    \mathbf{e}^{\textrm{RBF}}_{uv} = [\phi^\mathrm{e}_1(\tilde{X}^\mathrm{e}_{uv}), \phi^\mathrm{e}_2(\tilde{X}^\mathrm{e}_{uv}),  ...,  \phi^\mathrm{e}_{m_\mathrm{r}}(\tilde{X}^\mathrm{e}_{uv})],
\end{equation}
where $\tilde{\mathbf{X}}$ and $\tilde{\mathbf{X}}^\mathrm{e}$ are pre-normalized SAAO feature matrices, $\phi_{n}^\mathrm{h}(r) = \sin(\pi n r)$ is a sine function used for node (SAAO) embedding;
to improve the smoothess of the potential energy surface, we used the real Morlet wavelet functions for edge embedding:
\begin{equation}
    \phi_{m}^\mathrm{e}(r) = \exp(-(\frac{r}{\sigma\cdot c_\mathbf{X}})^2 ) \cdot \sin(\pi m r / c_\mathbf{X})
\end{equation}
and $c_\mathbf{X} \ (\mathbf{X}\in \set{\mathbf{F}, \mathbf{D}, \mathbf{P}, \mathbf{S}, \mathbf{H}})$ is the operator-specific upper cutoff value to $\tilde{X}^\mathrm{e}_{uv}$. To ensure size-consistency for energy predictions, a mollifier $I_\mathbf{X}(r)$ with the auxiliary edge attribute $ \mathbf{e}_{uv}^\textrm{aux}$ is introduced:
\begin{equation}
    \label{aux}
    \mathbf{e}_{uv}^{\textrm{aux}} = \mathbf{W}^{\textrm{aux}} \cdot I_\mathbf{X}(\tilde{X}^\mathrm{e}_{uv}) ,
\end{equation}
where
\begin{equation}
    \label{moll}
    I_\mathbf{X}(r) = 
     \begin{cases}
        \exp\left(\frac{c_\mathbf{X}}{|r|-c_\mathbf{X}} + 1\right) \cdot \exp(-(\frac{r}{\sigma\cdot c_\mathbf{X}})^2 ) \quad\text{if } 0 \le |r| < c_\mathbf{X} \\
       0 \quad\text{if } |r| \ge c_\mathbf{X} \\
     \end{cases}
\end{equation}
The radial basis function embeddings of the SAAOs and a one-hot encoding of the chemical element of the atoms ($\mathbf{f}_A^{\textrm{onehot}}$) are transformed by %
neural network modules to yield 0-th order SAAO, SAAO-pair, and atom attributes,
\begin{equation}
    \label{h0e0}
    \mathbf{h}_{u}^{0} = \mathrm{Enc}_\mathrm{h}(\mathbf{h}_{u}^{\textrm{RBF}}),\  \mathbf{e}_{uv}^{0} = \mathrm{Enc}_\mathrm{e}(\mathbf{e}_{uv}^{\textrm{RBF}}),\ 
    \mathbf{f}_{A}^{0} = \mathrm{Enc}_\mathrm{f}(\mathbf{f}_{A}^{\textrm{onehot}})
\end{equation}
where $\mathrm{Enc}_\mathrm{h}$ and $\mathrm{Enc}_\mathrm{e}$ are residual blocks\cite{resnet} comprising 3 dense NN layers, and $\mathrm{Enc}_\mathrm{f}$ is a single dense NN layer.
In contrast to atom-based message passing neural networks, this additional embedding transformation captures the interactions among the physical operators.

The update of the node- and edge-specific attributes (gray block in Fig.~\ref{fig:mpl}) is unchanged from Ref. ~\cite{orbnet1}, except with the additional information back-propagation from the atom-specific attributes. 
The node and edge attributes at step $l+1$ are updated via the following neural message passing mechanism (corresponding to ``AO-AO attention'' in Fig.~\ref{fig:mpl}):
\begin{subequations}\label{eq:mpl}
\begin{align}
    \tilde{\mathbf{h}}_u^{l+1} &= \mathbf{h}_{u}^l + \mathbf{W}_\mathrm{h,2}^{l} \cdot \textrm{Swish}\Big( \textrm{BatchNorm}\big(\mathbf{W}_\mathrm{h,1}^{l} \cdot \big[\bigoplus\limits_{i} (\sum_{v\in N(u)} w^{l,i}_{uv} \cdot \mathbf{m}_{uv}^l)\big] + \mathbf{b}_\mathrm{h,1}^{l}\big)\Big) + \mathbf{b}_\mathrm{h,2}^{l}\\
    \mathbf{m}_{uv}^{l} &= \textrm{Swish}(\mathbf{W}_\mathrm{m}^{l} \cdot [\mathbf{h}_u^l \odot \mathbf{h}_v^l \odot \mathbf{e}_{uv}^l] + \mathbf{b}_\mathrm{m}^{l})\\
    w^{l,i}_{uv} &= \textrm{Tanh}( \sum [(\mathbf{W}_\mathrm{a}^{l,i} \cdot \mathbf{h}_u^l) \odot (\mathbf{W}_\mathrm{a}^{l,i} \cdot \mathbf{h}_v^l) \odot \mathbf{e}_{uv}^l  \odot \mathbf{e}^{\textrm{aux}}_{uv}]/n_{\mathrm{e}})\\
    \mathbf{e}_{uv}^{l+1} &= \mathbf{e}_{uv}^{l} + \mathbf{W}_\mathrm{e,2}^{l} \cdot \big( \textrm{Swish}(\mathbf{W}_\mathrm{e,1}^{l} \cdot \mathbf{m}_{uv}^{l} + \mathbf{b}_\mathrm{e,1}^{l}) \big) + \mathbf{b}_\mathrm{e,2}^{l}
\end{align}
\end{subequations}

where $\mathbf{m}_{uv}^{l}$ is the message function on each edge,
$w^{l,i}_{uv}$, are multi-head attention scores \cite{gat} for the relative importance of SAAO pairs ($i$ indexes attention heads), $\bigoplus$ denotes a vector concatenation operation, $\odot$ denotes the Hadamard product, and $\cdot$ denotes the matrix-vector product.

The SAAO attributes are accumulated into the atoms on which the corresponding SAAOs are centered, using an attention-based pooling operation (``AO-Atom attention'' in Fig.~\ref{fig:mpl}) inspired by the set transformer \cite{lee2019set} architecture:
\begin{subequations}
\label{aoatomattn}
\begin{align}
    a_{A,u}^l &= \textrm{Softmax}(\mathbf{f}^l_A \cdot (\mathbf{h}_u^{l})^{\mathrm{T}}/\sqrt{n_{\mathrm{h}}}) \\
    \tilde{\mathbf{f}}_A^{l+1} &=  \mathbf{W}_\mathrm{f,1}^{l} \cdot \big[ \mathbf{f}_A^l || (\sum_{u\in A} a_{A,u}^l \mathbf{h}_{u}^l)\big]  + \mathbf{b}_\mathrm{f,1}^{l}
\end{align}    
\end{subequations}
where the Softmax operation is taken over all SAAOs $u$ centered on atom $A$. Then the global attention $\alpha_A^{l}$ is calculated for all atoms in the molecule to update the molecule-level attribute $\mathbf{q}^{l+1}$:
\begin{subequations}
\label{globattn}
\begin{align}
    \alpha_A^{l+1} &= \textrm{Softmax}(\mathbf{q}^l \cdot (\tilde{\mathbf{f}}_A^{l+1})^{\mathrm{T}}/\sqrt{n_{\mathrm{h}}}) \\
    \mathbf{q}^{l+1} &=  \mathbf{q}^{l} + \sum_{A} \alpha_{A}^{l+1} \tilde{\mathbf{f}}_{A}^{l+1}
\end{align}    
\end{subequations}
where the Softmax is taken over all atoms in the molecule, and the initial global attribute $\mathbf{q}^{0}$ is a molecule-independent, trainable parameter vector. 

Finally, the molecule- and atom-level information is propagated back to the SAAO attributes: 
\begin{subequations}
\begin{align}
    \mathbf{f}_A^{l+1} &=  \alpha_{A}^{l+1} \tilde{\mathbf{f}}_{A}^{l+1} \\
    \mathbf{h}_u^{l+1} &= \mathbf{W}_\mathrm{f,2}^{l} \cdot \big[ \mathbf{f}_{A}^{l+1} || \tilde{\mathbf{h}}_u^{l+1} \big]  + \mathbf{b}_\mathrm{f,2}^{l}.
\end{align}    
\end{subequations}

The list of trainable model parameters is: 
$\mathbf{W}^\mathrm{aux}$,
$\mathbf{W}^l_\mathrm{h,1}$,  $\mathbf{W}^l_\mathrm{h,2}$,  $\mathbf{b}^l_\mathrm{h,1}$,  $\mathbf{b}^l_\mathrm{h,2}$,  $\mathbf{W}^l_\mathrm{m}$, $\mathbf{b}^l_\mathrm{m}$, $\mathbf{W}^{l,i}_\mathrm{a}$, $\mathbf{W}^l_\mathrm{e,1}$, $\mathbf{W}^l_\mathrm{e,2}$, $\mathbf{b}^l_\mathrm{e,1}$, $\mathbf{b}^l_\mathrm{e,2}$, $\mathbf{W}^l_\mathrm{f,1}$, $\mathbf{W}^l_\mathrm{f,2}$, $\mathbf{b}^l_\mathrm{f,1}$, $\mathbf{b}^l_\mathrm{f,2}$, $\mathbf{q}^0$, and the parameters of $\mathrm{Enc}_\mathrm{h}$, $\mathrm{Enc}_\mathrm{e}$, $\mathrm{Enc}_\mathrm{f}$, $\mathrm{Dec}$, and $\mathrm{Dec}^\mathrm{aux}$.
%

%
%

\section{Model hyperparameters and training details}
\label{sec:hyperparams}
Table \ref{table:hp} summarizes the hyperparameters employed in this work. 
We perform a pre-transformation on the input features from $\mathbf{F}$, $\mathbf{D}$, $\mathbf{P}$, $\mathbf{H}$ and $\mathbf{S}$ to obtain $\tilde{\mathbf{X}}$ and $\tilde{\mathbf{X}}^\mathrm{e}$: We normalize all diagonal SAAO tensor values $X_{uu}$ to the range $[0, 1)$ for each operator type to obtain $\tilde{X}_{u}$; for off-diagonal SAAO tensor values, we take $\tilde{X}_{uv}=-\ln(|X_{uv}|)$ for $\mathbf{X} \in \Set{\mathbf{F}, \mathbf{P}, \mathbf{S}, \mathbf{H}}$, and $\tilde{D}_{uv}=D_{uv}$.

\begin{table*}[]
\centering
\setlength{\extrarowheight}{0.05cm}
\begin{tabular}{ccc}
\hline\hline
Hyperparameter    & Meaning   & Value  \\ \hline
$n_\mathrm{r}$   & Number of basis functions for node embedding & 8 \\ \hline
$m_\mathrm{r}$   & Number of basis functions for edge embedding & 8 \\ \hline
$n_\mathrm{h}$   & Dimension of hidden node attributes & 256 \\ \hline
$n_\mathrm{e}$   & Dimension of hidden edge attributes & 64  \\ \hline
$n_\mathrm{a}$   & Number of attention heads          & 4   \\ \hline
$L$              & Number of message passing \& pooling layers   & 2   \\ \hline
$L_\mathrm{enc}$ & Number of dense layers in $\mathrm{Enc}_\mathrm{h}$ and $\mathrm{Enc}_\mathrm{e}$ & 3   \\ \hline
$L_\mathrm{dec}$ & Number of residual blocks in a decoding network  & 3   \\ \hline
$n_\mathrm{d}$   & Hidden dimension of a decoding network & 256 \\ \hline
$\gamma$           & Batch normalization momentum  & 0.4 \\ \hline
$c_\mathbf{F}$     & Cutoff value for $\tilde{F}_{uv}$   & 6.0  \\ \hline
$c_\mathbf{D}$     & Cutoff value for $\tilde{D}_{uv}$ & 9.45  \\ \hline
$c_\mathbf{P}$     & Cutoff value for $\tilde{P}_{uv}$ & 6.0 \\ \hline
$c_\mathbf{S}$     & Cutoff value for $\tilde{S}_{uv}$ & 6.0  \\ \hline
$c_\mathbf{H}$     & Cutoff value for $\tilde{H}_{uv}$ & 6.0  \\ \hline
$\sigma$           & Morlet wavelet RBF scale     & 1/3  \\ \hline\hline
\end{tabular}
\caption{Model hyperparameters employed in OrbNet. All cutoff values are in atomic units.}
\label{table:hp}
\end{table*}

Training is performed on a loss function of the form
\begin{eqnarray}
    \mathcal{L}(\hat{\mathbf{E}}, \mathbf{E}, \hat{\mathbf{d}}, \mathbf{d}) &=& (1-\alpha) \sum_{i}\mathcal{L}_\mathrm{2}(\hat{E}_i, E_i)\nonumber\\
    &+& \alpha \sum_{i} \mathcal{L}_\mathrm{2}(\hat{E}_i-\hat{E}_{t(i)}, E_i-E_{t(i)})\\
    &+& \beta \sum_{i}\sum_{A\in i} \mathcal{L}_\mathrm{2}(\hat{\mathbf{d}}_A, \mathbf{d}_A).
    \label{eq:confloss}
\end{eqnarray}
$\sum_i$ denotes summation over a minibatch of molecular geometries $i$. For each geometry $i$, we randomly sample another conformer of the same molecule $t(i)$ to evaluate the relative conformer loss $\mathcal{L}_\mathrm{2}(\hat{E}_i-\hat{E}_{t(i)}, E_i-E_{t(i)})$; $\mathbf{E}$ denotes the ground truth energy values of the minibatch, $\hat{\mathbf{E}}$ denotes the model prediction values of the minibatch; $\hat{\mathbf{d}}_A$ and $\mathbf{d}_A$ denote the predicted and reference auxiliary target vectors for each atom $A$ in molecule $i$, and  $\mathcal{L}_\mathrm{2}(\hat{y},y)=||\hat{y}-y||_{2}^{2}$ denotes the L2 loss function. For the model used in %
Section \ref{sec:qm9}, we choose $\alpha=0$ as only the optimized geometries are available; for models in Section~\ref{sec:geo_opt}, we choose $\alpha=0.95$. $\beta$ is adaptively updated using the GradNorm\cite{chen2018gradnorm} method.

All models are trained on a single Nvidia Tesla V100-SXM2-32GB GPU using %
the Adam optimizer \cite{kingma2014adam}. For all training runs, we set the minibatch size to 64 and use a cosine annealing with warmup learning rate schedule \cite{goyal2017accurate} that performs a linear learning rate increase from $\num{3E-6}$ to $\num{3E-4}$ for the initial 100 epochs, and a cosine decay from $\num{3E-4}$ to $0$ for 200 epochs. 

\section{Analytical nuclear gradients for symmetry-adapted atomic-orbital features}

\label{sec:grad}

The electronic energy in the OrbNet model is given by
\begin{equation}
    E_{\text{out}}[\mathbf{f}] = E_{\text{xTB}} + E_{\text{NN}} [\mathbf{f}].
\end{equation}
Here, $\mathbf{f}$ denotes the features, which correspond to the matrix elements of the quantum mechanical operators $\{\mathbf{F}, \mathbf{P}, \mathbf{D}, \mathbf{H}, \mathbf{S}\}$ evaluated in the SAAO basis. 

\subsection{Generation of SAAOs}
\label{SAAO}
We denote $\{\phi^{A}_{n,l,m}\}$ as the set of atomic basis functions with atom indices $A$, with principle, angular and magnetic quantum numbers $n, l, m$, and $\{\psi_i\}$ as the set of canonical molecular orbitals obtained from a low-level electronic structure calculation.

We define the transformation matrix $\mathbf{X}$ between AOs and SAAOs as eigenvectors of the local density matrices (in covariant form):
\begin{align} \label{eq:saao_eigen}
\tilde{\mathbf{P}}^{A}_{n,l}  \mathbf{X}^{A}_{n,l} = \mathbf{X}^{A}_{n,l} \Sigma^{A}_{n,l} 
\end{align}
where $\tilde{\mathbf{P}}$ is the covariant density matrix in AO basis and is defined as
\begin{align}
\tilde{\mathbf{P}} = \mathbf S \mathbf P^\text{AO} \mathbf S
\end{align}

The SAAOs, $\{\hat \phi_\kappa\}$, are thus expressed as
\begin{equation} \label{eq:saao_transform}
     |\hat \phi_{\kappa} \rangle = \sum_{\mu} X_{\mu \kappa} |\phi_{\mu} \rangle
\end{equation}

\subsection{Matrices of operators in the SAAO basis for featurization}

\begin{itemize}

\item The xTB core-Hamiltonian matrix in the SAAO basis
\begin{equation}
    \mathbf H^\text{SAAO} =  \mathbf X^\dagger \mathbf H^\text{AO} \mathbf X
\end{equation}
\item Overlap matrix in the SAAO basis
\begin{equation}
    \mathbf S^\text{SAAO} =  \mathbf X^\dagger \mathbf S^\text{AO} \mathbf X
\end{equation}
\item The xTB Fock matrix in the SAAO basis
\begin{equation}
    \mathbf F^\text{SAAO} =  \mathbf X^\dagger \mathbf F^\text{AO} \mathbf X
\end{equation}
\item Density matrix in the SAAO basis
\begin{equation}
    \mathbf P^\text{SAAO} =  \mathbf X^\dagger \mathbf P^\text{AO} \mathbf X
\end{equation}
\item Centroid distance matrix in the SAAO basis
\begin{align}
D^\text{SAAO}_{\kappa \lambda} = || \langle \hat{\phi}_{\kappa} | \hat{\mathbf{r}} | \hat{\phi}_{\kappa} \rangle - \langle \hat{\phi}_{\lambda} | \hat{\mathbf{r}} | \hat{\phi}_{\lambda} \rangle || = \left( \vec d_{\kappa \lambda} \cdot   \vec d_{\kappa \lambda} \right)^{1/2}
\end{align}
where $\vec d_{\kappa \lambda}$ is defined as
\begin{align}
\vec d_{\kappa \lambda} = \vec r^\text{SAAO}_{\kappa \kappa} - \vec r^\text{SAAO}_{\lambda \lambda} =  \left( \mathbf X^\dagger \mathbf r^\text{AO} \mathbf X \right)_{\kappa\kappa} - \left( \mathbf X^\dagger \mathbf r^\text{AO} \mathbf X \right)_{\lambda\lambda}
\end{align}
where $\mathbf r^\text{AO}$ is the AO dipole matrix.

\end{itemize}

\subsection{OrbNet analytical gradient}

The Lagrangian for OrbNet is 
\begin{align}
    \mathcal{L} & = E_{\text{NN}} [\mathbf{f}]+ \sum_{pq} W_{pq} \left( \mathbf{C}^{\dagger} \mathbf{S} \mathbf{C} - \mathbf{I} \right)_{pq}  + \sum_{ai} z_{ai} F_{ai} \label{eq:L}
\end{align}
Second term: orbitals orthogonality constraint.
Third term: Brillion conditions.
Note: $i,j$ are indices for occupied molecular orbitals (MOs), $p,q$ are general indices for MOs.

\subsection{Stationary condition for the Lagrangian with respect to the MOs}

The Lagrangian is stationary with respect to variations of the MOs:
\begin{align} \label{eq:stationary_cond}
   \frac{\partial \mathcal{L}}{\partial V_{pq}} = 0
\end{align}
where $V_{pq}$ is a variation of the MOs in terms of the orbital rotation between MO pair $p$ and $q$ and is defined as
\begin{align}
\tilde{\mathbf C} = \mathbf C (\mathbf I + \mathbf V)
\end{align}

This leads to the following expressions for each term on the right-hand-side of Eq. \ref{eq:L}:
\begin{align}
A_{pq} &= \frac{\partial E_{\text{NN}}[\mathbf f]}{\partial V_{pq}} \Bigr\vert_{\mathbf V = 0}  = \frac{\partial E_{\text{NN}}[\mathbf f]}{\partial \mathbf f} \frac{\partial \mathbf f}{\partial V_{pq}} \Bigr\vert_{\mathbf V = 0}  \label{eq:Apq}  \\
W_{pq} &= \frac{\partial \sum_{pq} W_{pq} \paran*{ \mathbf{C}^{\dagger} \mathbf{S} \mathbf{C} - \mathbf{I} }_{pq} }{\partial V_{pq}} \Bigr\vert_{\mathbf V = 0}   \\
(\mathbf A[\mathbf z])_{pq} &= \frac{\partial \sum_{ai} z_{ai} F_{ai} }{\partial V_{pq}} \Bigr\vert_{\mathbf V = 0} = \left( \mathbf{F} \mathbf{z} \right)_{p q} \Big|_{q \in \text{occ}} + \left( \mathbf{F} \mathbf{z}^{\dagger} \right)_{p q} \Big|_{q \in \text{vir} } + 2 \left( \mathbf{g}[ \bar{\mathbf{z}} ] \right)_{p q} \Big{|}_{q \in \text{occ}}
\end{align}

In the following sections, we derive the working equations for the above terms.

\subsection{SAAO derivatives}

As will be shown later, the OrbNet energy gradient involves the derivatives of the SAAO transformation matrix $\mathbf{X}^{A}_{n,l}$ with respect to orbital rotations and nuclear coordinates. The derivatives of SAAOs are a bit involved, since SAAOs are eigenvectors of the local density matrices. We follow reference \cite{magnus1985} and show how SAAO derivatives are computed.

Here, we restrict the discussion to the scenario where the eigenvalues of the local density matrices $\tilde{\mathbf{P}}^{A}_{n,l}$ are distinct, such that the eigenvectors (i.e. SAAOs) are uniquely determined up to a constant (real-valued) factor.

For generality, denote the (real, symmetric) matrix for which the eigenvalues/eigenvectors are solved as $\mathbf A$, its eigenvalues as $\boldsymbol{\Lambda}$, and its eigenvectors as $\mathbf X$, such that
\begin{align}
\mathbf A \mathbf X = \mathbf X \boldsymbol \Lambda
\end{align}
with the eigenvectors $\mathbf X$ being orthonormal to each other,
\begin{align}
\mathbf X^T \mathbf X =  \mathbf I
\end{align}

Denote the derivative of a martrix with respect to a parameter $p$ by a prime, for example,
\begin{align}
\frac{d \mathbf A}{d p} \equiv \mathbf A'
\end{align}

The eigenvalue derivatives are computed as
\begin{align}
\frac{d \lambda_k}{d p} = {\mathbf X}_k^\dagger \mathbf A' {\mathbf X}_k
\end{align}

Define matrix $\mathbf T$ as
\begin{align} \label{eq:T-a}
\mathbf T = \mathbf X^{-1} \mathbf X'
\end{align}

For the case where the eigenvalues are distinct, we have
\begin{align}
T_{kl} &= \frac{{\mathbf X}_k^\dagger \mathbf A' {\mathbf X}_l}{\lambda_l - \lambda_k} \quad \text{for } k \neq l, \quad
T_{kk} = 0
\label{eq:T-b}
\end{align}

The eigenvector derivative can be determined via \cref{eq:T-a}, as
\begin{align} \label{eq:X-der}
\mathbf X' = \mathbf X \mathbf T
\end{align}

Let's denote a diagonal block of the covariant density matrix on atom A with quantum numbers $\{n, l\}$ as $I$, such that
\begin{align}
\tilde{\mathbf{P}}_I \equiv \tilde{\mathbf{P}}^{A}_{n,l}
\end{align}

The SAAO eigenvalue problem for the $I$-th diagonal block can thus be re-written as
\begin{align}
\tilde{\mathbf{P}}_I \mathbf X_I = \mathbf X_I \boldsymbol{\Sigma}_I
\end{align}

The derivatives of  $\mathbf X_I$ with respect to an arbitrary variable $\xi$, denoted as $\mathbf X_I^\xi$, can be expressed as:
\begin{align}
\mathbf X_I^\xi = \mathbf X_I \mathbf T_I^\xi
\end{align}
where matrix $\mathbf T_I^\xi$ is defined according to \cref{eq:T-b} as
\begin{align}
T_{I, \kappa\lambda}^\xi &= \frac{{\mathbf X}_{I, \kappa}^\dagger \tilde{\mathbf{P}}_I^\xi {\mathbf X}_{I, \lambda}}{\epsilon_\lambda - \epsilon_\kappa} \quad \text{for } \kappa \neq \lambda, \quad
T_{I, \kappa\kappa}^\xi = 0
\label{eq:T-xi}
\end{align}
where $\tilde{\mathbf{P}}_I^\xi$ is the derivative of the covariant density matrix for $I$-th diagonal block.

The gradients of the OrbNet energy usually involve the term $A^{\xi} \equiv \text{Tr} [\mathbf B \mathbf T^{\xi}] $, 
which can be re-written as
\begin{align}
\text{Tr} [\mathbf B \mathbf T^{\xi}]  &= \sum_I \text{Tr} [\mathbf B_I \mathbf T^{\xi}_I] \notag 
= \sum_I \sum_{\kappa \neq \lambda} B_{I, \kappa\lambda} T^{\xi}_{I, \lambda\kappa} \notag 
= \sum_I \sum_{\kappa \neq \lambda} B_{I, \kappa\lambda}  \frac{{\mathbf X}_{I, \lambda}^\dagger \tilde{\mathbf{P}}_I^\xi {\mathbf X}_{I, \kappa}}{\epsilon_\kappa - \epsilon_\lambda}
\end{align}

Define $\bar{\mathbf B}_I$ as
\begin{align}
\bar{B}_{I, \kappa\lambda}  &= \frac{B_{I, \kappa\lambda}}{\epsilon_\kappa - \epsilon_\lambda} \quad \text{for } \kappa \neq \lambda, \quad
\bar{B}_{I, \kappa\kappa}  = 0
\label{eq:B-bar}
\end{align}
and $\tilde{\mathbf B}_I$ as
\begin{align} 
\tilde{\mathbf B}_I  = \frac{1}{2} {\mathbf X}_{I} (\bar{\mathbf B}_I + \bar{\mathbf B}_I^\dagger) {\mathbf X}_{I}^\dagger
\label{eq:B-tilde}
\end{align}

where we have symmetrized $\tilde{\mathbf B}_I$. Finally, we have
\begin{equation} \label{eq:TrBT}
\text{Tr} [\mathbf B \mathbf T^{\xi}] = \sum_I \sum_{\kappa \lambda} \bar{B}_{I, \kappa\lambda}  {\mathbf X}_{I, \lambda}^\dagger \tilde{\mathbf{P}}_I^\xi {\mathbf X}_{I, \kappa} \notag 
= \sum_I \text{Tr} [\tilde{\mathbf B}_{I} \tilde{\mathbf{P}}_I^\xi ]
\end{equation}

\subsubsection{Derivatives of the SAAO basis with respect to MO variations}

The derivatives of the SAAOs, $\mathbf X$ with respect to orbital variation $V_{pq}$ can be expressed as:
\begin{align}
\frac{\partial \mathbf X}{\partial V_{pq}} =  \mathbf X \mathbf T^{V_{pq}} \label{eq:dX_orb}
\end{align}
where $\mathbf T^{V_{pq}}$ is defined according to \cref{eq:T-xi} as
\begin{align}
T_{I, \kappa\lambda}^{V_{pq}} &= \frac{\mathbf X_{I,\kappa}^T \tilde{\mathbf{P}}_I^{V_{pq}} \mathbf X_{I,\lambda} } {\epsilon_{I,\lambda} - \epsilon_{I,\kappa}} \quad \text{for } \kappa \neq \lambda , \quad
T_{I, \kappa\kappa}^{V_{pq}}  = 0
\label{eq:T-pq}
\end{align}
where $\tilde{\mathbf{P}}_I^{V_{pq}}$ is the derivative of the $I$-th diagonal block of local density matrix with respect to orbital variation $V_{pq}$ and is defined as
\begin{align}
\tilde P_{\mu\nu \in I}^{V_{pq}} &\equiv \frac{\partial \tilde P^I_{\mu\nu}}{\partial V_{pq}} = \frac{\partial (\mathbf S \mathbf P \mathbf S)_{\mu\nu}}{\partial V_{pq}} \notag 
=  \sum_{\kappa \lambda} S_{\mu\kappa}  \frac{\partial P_{\kappa \lambda}}{\partial V_{pq}} S_{\lambda\nu} \notag \\
&= \sum_{\kappa \lambda} S_{\mu\kappa} (C_{\kappa p} n_q C_{\lambda q} + C_{\kappa q} n_q C_{\lambda p}) S_{\lambda\nu} 
\end{align}
where $n_q$ is the occupation number of orbital $q$. For closed-shell systems at zero electronic temperature, $n_q$ is defined as
\begin{align}
n_q =  \begin{cases}
2 & \mbox{if } q \in \mbox{ occupied} \\
0 & \text{otherwise}
\end{cases}
\end{align}
For other cases, $n_q$ may be fractional numbers.

Define $\mathbf{Y} = \mathbf S \mathbf C$,
then
\begin{align}
\tilde{\mathbf P}_{I}^{V_{pq}} &= (\mathbf Y_p \mathbf Y^\dagger_q + \mathbf Y_q \mathbf Y^\dagger_p) n_q
\end{align}

The orbital derivatives of the OrbNet energy usually involve the term $\text{Tr} [\mathbf B \mathbf T^{pq}]$, 
which can be expressed according to \cref{eq:TrBT}
as
\begin{align}
\text{Tr} [\mathbf B \mathbf T^{pq}]  &= \sum_I \text{Tr} [\tilde{\mathbf B}_I \tilde{\mathbf{P}}_I^{V_{pq}} ] \notag = \sum_I \text{Tr} [\tilde{\mathbf B}_I (\mathbf Y_p \mathbf Y^\dagger_q + \mathbf Y_q \mathbf Y^\dagger_p) n_q ] \notag = M_{pq} 
\end{align}
where $\tilde{\mathbf B}_I$ is defined in \cref{eq:B-tilde}; $M_{pq}$ is defined as
\begin{align} \label{eq:M}
M_{pq} = 2 \mathbf Y_p^\dagger \tilde{\mathbf B} \mathbf Y_q n_q
\end{align}

\subsubsection{Derivatives of the SAAO basis with respect to nuclear coordinates}

The derivatives of $\mathbf X$ with respect to nuclear coordinates $x$ can be expressed as
\begin{align}
\frac{\partial \mathbf X}{\partial x} =  \mathbf X  \mathbf T^{x} \label{eq:dX_nuc}
\end{align}
where $\mathbf T^{x}$ is defined according to \cref{eq:T-xi} as
\begin{align}
T_{I, \kappa\lambda}^{x} &= \frac{\mathbf X_{I,\kappa}^T \tilde{\mathbf{P}}_I^{x} \mathbf X_{I,\lambda} } {\epsilon_{I,\lambda} - \epsilon_{I,\kappa}} \quad \text{for } \kappa \neq \lambda , \quad
T_{I, \kappa\kappa}^{x}  = 0
 \label{eq:T-pq}
\end{align}

where $\tilde{\mathbf{P}}_I^x$ is defined as
\begin{align}
\tilde P_{\mu\nu \in I}^{x} &\equiv \frac{\partial \tilde P^I_{\mu\nu}}{\partial x} = \frac{\partial (\mathbf S \mathbf P \mathbf S)_{\mu\nu}}{\partial x} \notag \\
&=  \sum_{\kappa \lambda} \frac{\partial S_{\mu\kappa}}{\partial x}  P_{\kappa \lambda} S_{\lambda\nu} + S_{\mu\kappa} P_{\kappa \lambda} \frac{\partial S_{\lambda\nu}}{\partial x} \notag \\
&= \sum_{\kappa \lambda} S_{\mu\kappa}^x P_{\kappa \lambda} S_{\lambda\nu}  + S_{\mu\kappa} P_{\kappa \lambda} S_{\lambda\nu}^x
\end{align}

Define $N = \mathbf P \mathbf S$,
then
\begin{align}
\tilde{\mathbf P}_I^x = \left[ \mathbf S^x \mathbf N + \mathbf N^\dagger \mathbf S^x \right]_I
\end{align}

The nuclear derivatives of the OrbNet energy usually involve the term $\text{Tr} [\mathbf B \mathbf T^{x}] $,
which can be re-written according to Eq. \ref{eq:TrBT} as
\begin{align}
\text{Tr} [\mathbf B \mathbf T^{x}]  &= \sum_I \text{Tr} [\tilde{\mathbf B}_I \tilde{\mathbf{P}}_I^{x} ] \notag = \sum_I \text{Tr} \left( \tilde{\mathbf B}_I \left[ \mathbf S^x \mathbf N + \mathbf N^\dagger \mathbf S^x \right]_I \right) \notag =  \text{Tr} [\overline{\mathbf W} \mathbf S^x]
\end{align}
where $\tilde{\mathbf B}$ defined in Eq. \ref{eq:B-tilde}; $\overline{\mathbf W}$ is defined as
\begin{align} \label{eq:W-bar}
\overline{W}_{\mu\nu}^I \Bigr \vert_{\nu \in I}= 2 \left( \mathbf N \tilde{\mathbf B}_I \right)_{\mu\nu}
\end{align}

\subsection{Derivatives of OrbNet energy with respect to the MOs}

Define the derivatives of the OrbNet energy with respect to feature $\mathbf f$ as:
\begin{align} \label{eq:dE_df}
\mathbf Q^f = \frac{\partial E_{\text{NN}}[\mathbf f]}{\partial \mathbf f}
\end{align}
where $\mathbf f \in \{\mathbf{F}, \mathbf{P}, \mathbf{D}, \mathbf{H}, \mathbf{S}\}$.

Note that $\mathbf Q^f$ has the same dimension as $\mathbf f$, and is 
 symmetrized.

The derivatives of OrbNet energy with respect to the MO variations, Eq. \ref{eq:Apq},  can be rewritten as
\begin{align}
A_{pq} &= \frac{\partial E_{\text{NN}}[\mathbf f]}{\partial V_{pq}} \Bigr\vert_{\mathbf V = 0}  = \frac{\partial E_{\text{NN}}[\mathbf f]}{\partial \mathbf f} \frac{\partial \mathbf f}{\partial V_{pq}} \Bigr\vert_{\mathbf V = 0}  = \sum_f \left[ \mathbf Q^f \cdot \frac{\partial \mathbf f}{\partial V_{pq}} \right]
\end{align}
Define
\begin{align}
A_{pq}^f = \mathbf Q^f \cdot \frac{\partial \mathbf f}{\partial V_{pq}} 
\end{align}
which corresponds to the contribution to OrbNet energy derivatives with respect to MOs from a specific feature $\mathbf f$.
We then derive the expression of $A_{pq}^f$ for each individual feature, as described below.

\subsubsection{Core Hamiltonian}

\begin{align} 
A_{pq}^H &= \mathbf Q^H  \cdot \frac{\partial \mathbf H^\text{SAAO}}{\partial V_{pq}} =  \mathbf Q^H  \cdot \left[ \mathbf X^\dagger \mathbf H^\text{AO} \frac{\partial \mathbf X}{\partial V_{pq}} + \text{transpose} \right] \notag \\
&= \mathbf Q^H  \cdot \left[ \mathbf X^\dagger \mathbf H^\text{AO} \mathbf X \mathbf T^{pq} + \text{transpose} \right] \notag \\
&= 2 \text{Tr} \left[ \mathbf Q^H   \mathbf H^\text{SAAO} \mathbf T^{pq} \right] \notag \\
&= 2 \text{Tr} \left[ \mathbf B^H \mathbf T^{pq} \right] \notag \\
&= 2 M^H_{pq}
\end{align}
where $M^H_{pq}$ is defined according to Eq. \ref{eq:M}:
\begin{align} \label{eq:M-H}
M^H_{pq} =  \mathbf Y_p^\dagger (\tilde{\mathbf B}^H +  \tilde{\mathbf B}^{H, \dagger}) \mathbf Y_q n_q
\end{align}
and $\mathbf B^H$ is
\begin{align}
\mathbf B^H = \mathbf Q^H   \mathbf H^\text{SAAO} 
\end{align}

\subsubsection{Overlap matrix}

\begin{align} 
A_{pq}^S &= \mathbf Q^S  \cdot \frac{\partial \mathbf S^\text{SAAO}}{\partial V_{pq}} = \mathbf Q^S  \cdot \left[  \mathbf X^\dagger \mathbf S^\text{AO} \frac{\partial \mathbf X}{\partial V_{pq}} + \text{transpose} \right] \notag \\
&= 2 \text{Tr} \left[ \mathbf Q^S  \mathbf S^\text{SAAO} \mathbf T^{pq} \right] \notag \\
&= 2 \text{Tr} \left[ \mathbf B^S \mathbf T^{pq} \right] \notag \\
&= 2 M^S_{pq}
\end{align}
with $M^S_{pq}$ defined in a similar way as to Eq. \ref{eq:M-H}.

\subsubsection{Fock matrix}

\begin{align} 
A_{pq}^F &= \mathbf Q^F  \cdot \frac{\partial \mathbf F^\text{SAAO}}{\partial V_{pq}}
\end{align}
with
\begin{align} 
\frac{\partial \mathbf F^\text{SAAO}}{\partial V_{pq}} &= \left( \mathbf X^\dagger \mathbf F^\text{AO} \frac{\partial \mathbf X}{\partial V_{pq}} + \text{transpose} \right) + \mathbf X^\dagger \frac{\partial \mathbf F^\text{AO}}{\partial V_{pq}} \mathbf X 
\end{align}
Therefore,
\begin{align} 
A_{pq}^F &= 2 \text{Tr} \left[ \mathbf Q^F   \mathbf F^\text{SAAO} \mathbf T^{pq} \right] + \text{Tr} \left[ \mathbf Q^F \mathbf X^\dagger \frac{\partial \mathbf F^\text{AO}}{\partial V_{pq}} \mathbf X \right] \notag \\
&= 2 \text{Tr} \left[ \mathbf B^F \mathbf T^{pq} \right] + \text{Tr} \left[ \mathbf Q^{F, \text{AO}} \frac{\partial \mathbf F^\text{AO}}{\partial V_{pq}} \right] \notag \\
&= 2 \text{Tr} \left[ \mathbf B^F \mathbf T^{pq} \right] + \text{Tr} \left[ \mathbf Q^{F, \text{AO}} \frac{\partial \mathbf F^\text{AO}}{\partial \mathbf D^\text{AO}}  \frac{\partial \mathbf D^\text{AO}}{\partial V_{pq}} \right] \notag \\
&= 2 \text{Tr} \left[ \mathbf B^F \mathbf T^{pq} \right] + 2 \left( \mathbf g [2\mathbf Q^{F, \text{AO}}] \right)_{pq} \notag \\
&= 2 \text{Tr} \left[ \mathbf B^F \mathbf T^{pq} \right] + 2 \left( \mathbf g [2\mathbf Q^{F, \text{AO}}] \right)_{pq} \notag \\
&= 2 M^F_{pq} + 2 \left( \mathbf g [2\mathbf Q^{F, \text{AO}}] \right)_{pq}
\end{align}
where $\mathbf g$ is the generalized xTB Fock matrix and $\mathbf Q^{F, \text{AO}}$ is defined as
\begin{align}
\mathbf Q^{F, \text{AO}} = \mathbf X \mathbf Q^F \mathbf X^\dagger
\end{align}

\subsubsection{Density matrix}

\begin{align} 
A_{pq}^P &= \mathbf Q^P  \cdot \frac{\partial \mathbf P^\text{SAAO}}{\partial V_{pq}}
\end{align}
where
\begin{align} 
\frac{\partial \mathbf P^\text{SAAO}}{\partial V_{pq}} = \left( \mathbf X^\dagger \mathbf P^\text{AO} \frac{\partial \mathbf X}{\partial V_{pq}} + \text{transpose} \right) + \mathbf X^\dagger \frac{\partial \mathbf P^\text{AO}}{\partial V_{pq}} \mathbf X
\end{align}
Therefore,
\begin{align}
A_{pq}^P &= 2 \text{Tr} \left[ \mathbf Q^P   \mathbf P^\text{SAAO} \mathbf T^{pq} \right] + 2 \left(\mathbf C^\dagger \mathbf Q^{P, \text{AO}} \mathbf C \right)_{pq} n_q \notag \\
&= 2 \text{Tr} \left[ \mathbf B^P T^{pq} \right] + 2 \left(\mathbf C^\dagger \mathbf Q^{P, \text{AO}} \mathbf C \right)_{pq} n_q \notag \\
&= 2 M^P_{pq} + 2 \left(\mathbf C^\dagger \mathbf Q^{P, \text{AO}} \mathbf C \right)_{pq} n_q
\end{align}
 
\subsubsection{Centroid distance matrix}

\begin{align} 
A_{pq}^D &= \mathbf Q^D  \cdot \frac{\partial \mathbf D^\text{SAAO}}{\partial V_{pq}}
\end{align}
with
\begin{align}
\frac{\partial D^\text{SAAO}_{\kappa \lambda}}{\partial V_{pq}} = \frac{1}{D^\text{SAAO}_{\kappa \lambda}} \vec{d}_{\kappa \lambda} \cdot \frac{\partial \vec{d}_{\kappa \lambda}}{\partial V_{pq}}
\end{align}
where $\frac{\partial \vec{d}_{\kappa \lambda}}{\partial V_{pq}}$ is defined as
\begin{align}
\frac{\partial \vec{d}_{\kappa \lambda}}{\partial V_{pq}} &=  \left( \mathbf X_\kappa^\dagger \mathbf r^\text{AO} \frac{\partial \mathbf X_\kappa}{\partial V_{pq}} - \mathbf X_\lambda^\dagger \mathbf r^\text{AO} \frac{\partial \mathbf X_\lambda}{\partial V_{pq}} \right) + \text{transpose} \notag \\
&= 2 \left[ (\mathbf r^\text{SAAO} \mathbf T^{pq})_{\kappa\kappa} - (\mathbf r^\text{SAAO} \mathbf T^{pq})_{\lambda\lambda}\right]
\end{align}

Then
\begin{align} 
A_{pq}^D &= \mathbf Q^D  \cdot \frac{\partial \mathbf D^\text{SAAO}}{\partial V_{pq}} \notag \\
&= \sum_{\kappa \lambda}  \frac{Q^D_{\kappa \lambda}}{D^\text{SAAO}_{\kappa \lambda}} \vec d_{\kappa \lambda} \cdot \frac{\partial \vec d_{\kappa \lambda}}{\partial V_{pq}}
\end{align}

Define
\begin{align} 
\vec R_{\kappa \lambda} &=  \frac{Q^D_{\kappa \lambda} }{D^\text{SAAO}_{\kappa \lambda}} \vec d_{\kappa \lambda} \label{eq:R} 
\end{align}

Then
\begin{align} 
A_{pq}^D
&= \sum_{\kappa \lambda}  \vec R_{\kappa \lambda} \cdot \frac{\partial \vec d_{\kappa \lambda}}{\partial V_{pq}} \notag \\
&= 2 \sum_{\kappa \lambda}  \vec R_{\kappa \lambda} \cdot  \left[ (\mathbf r^\text{SAAO} \mathbf T^{pq})_{\kappa\kappa} - (\mathbf r^\text{SAAO} \mathbf T^{pq})_{\lambda\lambda}\right] \notag \\
&= 4 \sum_{\kappa \lambda}  \vec R_{\kappa \lambda} \cdot  (\mathbf r^\text{SAAO} \mathbf T^{pq})_{\kappa\kappa} 
\end{align}

Define
\begin{align} 
\vec R^L_{\kappa} &= \sum_\lambda \vec R_{\kappa \lambda} \label{eq:R-L}  \\
B^D_{\kappa \lambda} &= \vec R^L_{\kappa} \cdot \vec r^\text{SAAO}_{\kappa \lambda}
\end{align}

Then
\begin{align} 
A_{pq}^D
&= 4 \text{Tr} \left[ \mathbf B^D \mathbf T^{pq} \right] \notag \\
&= 4 M^S_{pq}
\end{align}
where $M^D_{pq}$ defined in a similar way as to Eq. \ref{eq:M-H}.

\subsection{Derivatives of OrbNet energy with respect to nuclear coordinates}

The derivatives of OrbNet energy with respect to nuclear coordinates can be written as
\begin{align}
\frac{\partial E_{\text{NN}}}{\partial x} &= \frac{\partial E_{\text{NN}}[\mathbf f]}{\partial \mathbf f} \frac{\partial \mathbf f}{\partial x} = \sum_f \left[ \mathbf Q^f \frac{\partial \mathbf f}{\partial x} \right]
\end{align}

Define:
\begin{align}
A^f_x = \mathbf Q^f \cdot \frac{\partial \mathbf f}{\partial x} 
\end{align}
which corresponds to the contribution to OrbNet energy derivatives with respect to MOs from a specific feature $\mathbf f$.
x
Now let's derive the expression of $A_{x}^f$ for each individual feature:

\subsubsection{Core Hamiltonian}

\begin{align} 
A_{x}^H &= \mathbf Q^H  \cdot \frac{\partial \mathbf H^\text{SAAO}}{\partial x} =  \mathbf Q^H  \cdot \left[ \mathbf X^\dagger \mathbf H^\text{AO} \frac{\partial \mathbf X}{\partial x} + \text{transpose} \right] \notag \\
&= \mathbf Q^H  \cdot \left[ \mathbf X^\dagger \mathbf H^\text{AO} \mathbf X \mathbf T^{x} + \text{transpose} \right] \notag \\
&= 2 \text{Tr} \left[ \mathbf Q^H   \mathbf H^\text{SAAO} \mathbf T^{x} \right] + \text{Tr} \left[ \mathbf Q^H \mathbf X^\dagger \frac{\partial \mathbf H^\text{AO}}{\partial x} \mathbf X \right] \notag \\
&= 2 \text{Tr} \left[ \mathbf Q^H   \mathbf H^\text{SAAO} \mathbf T^{x} \right] + \text{Tr} \left[ \mathbf Q^{H,\text{AO}} \frac{\partial \mathbf H^\text{AO}}{\partial x} \right] \notag \\
&= 2 \text{Tr} \left[ \mathbf B^H  \mathbf T^{x} \right] + \text{Tr} \left[ \mathbf Q^{H,\text{AO}} \frac{\partial \mathbf H^\text{AO}}{\partial x} \right] \notag \\
&= 2 \text{Tr} \left[ \overline{\mathbf W}^H \mathbf S^x \right] + \text{Tr} \left[ \mathbf Q^{H,\text{AO}} \frac{\partial \mathbf H^\text{AO}}{\partial x} \right]
\end{align}
where $\overline{\mathbf W}^H$ is defined according to Eq. \ref{eq:W-bar}.

\subsubsection{Overlap matrix}

\begin{align} 
A_{x}^S &= \mathbf Q^S  \cdot \frac{\partial \mathbf S^\text{SAAO}}{\partial x} = \mathbf Q^S  \cdot \left[  \mathbf X^\dagger \mathbf S^\text{AO} \frac{\partial \mathbf X}{\partial x} + \text{transpose} \right] \notag \\
&= 2 \text{Tr} \left[ \mathbf Q^S  \mathbf S^\text{SAAO} \mathbf T^{x} \right] + \text{Tr} \left[ \mathbf Q^S \mathbf X^\dagger \frac{\partial \mathbf S^\text{AO}}{\partial x} \mathbf X \right] \notag \\
&= 2 \text{Tr} \left[ \mathbf Q^S  \mathbf S^\text{SAAO} \mathbf T^{x} \right] + \text{Tr} \left[ \mathbf Q^{S, \text{AO}}  \frac{\partial \mathbf S^\text{AO}}{\partial x} \right] \notag \\
&= 2 \text{Tr} \left[ \mathbf B^S \mathbf T^{x} \right] + \text{Tr} \left[ \mathbf Q^{S, \text{AO}}  \frac{\partial \mathbf S^\text{AO}}{\partial x} \right] \notag \\
&= 2 \text{Tr} \left[ \overline{\mathbf W}^S \mathbf S^x \right] + \text{Tr} \left[ \mathbf Q^{S, \text{AO}} \mathbf S^x  \right] \notag \\
&= \text{Tr} \left[ ( \overline{\mathbf W}^S + \mathbf Q^{S, \text{AO}} )  \mathbf S^x  \right] 
\end{align}
where $\overline{\mathbf W}^S$ is defined according to Eq. \ref{eq:W-bar}.

\subsubsection{Fock matrix}

\begin{align} 
A_{x}^F &= \mathbf Q^F  \cdot \frac{\partial \mathbf F^\text{SAAO}}{\partial x}
\end{align}
with
\begin{align} 
\frac{\partial \mathbf F^\text{SAAO}}{\partial x} &=\left( \mathbf X^\dagger \mathbf F^\text{AO} \frac{\partial \mathbf X}{\partial x} + \text{transpose} \right)+ \mathbf X^\dagger \frac{\partial \mathbf F^\text{AO}}{\partial x} \mathbf X 
\end{align}
Therefore,
\begin{align} 
A_{x}^F &= 2 \text{Tr} \left[ \mathbf Q^F   \mathbf F^\text{SAAO} \mathbf T^{x} \right] + \text{Tr} \left[ \mathbf Q^F \mathbf X^\dagger \frac{\partial \mathbf F^\text{AO}}{\partial x} \mathbf X \right] \notag \\
&= 2 \text{Tr} \left[ \mathbf Q^F   \mathbf F^\text{SAAO} \mathbf T^{x} \right] + \text{Tr} \left[ \mathbf Q^{F, \text{AO}} \frac{\partial \mathbf F^\text{AO}}{\partial x} \right] \notag \\
&=  2 \text{Tr} \left[ \mathbf B^F \mathbf T^{x} \right] + \text{Tr} \left[ \mathbf Q^{F, \text{AO}} \frac{\partial \mathbf F^\text{AO}}{\partial x} \right] \notag \\
&= 2 \text{Tr} \left[ \overline{\mathbf W}^F \mathbf S^x \right]  + \text{Tr} \left[ \mathbf Q^{F, \text{AO}} \frac{\partial \mathbf F^\text{AO}}{\partial x} \right]
\end{align}
where $\overline{\mathbf W}^F$ is defined according to Eq. \ref{eq:W-bar}.

\subsubsection{Density matrix}
\begin{align} 
A_{x}^P &= \mathbf Q^P  \cdot \frac{\partial \mathbf P^\text{SAAO}}{\partial x}
\end{align}
where
\begin{align} 
\frac{\partial \mathbf P^\text{SAAO}}{\partial x} = \mathbf X^\dagger \mathbf P^\text{AO} \frac{\partial \mathbf X}{\partial x} + \text{transpose} 
\end{align}
Therefore,
\begin{align}
A_{x}^P &= 2 \text{Tr} \left[ \mathbf Q^P   \mathbf P^\text{SAAO} \mathbf T^{x} \right] \notag \\
&= 2 \text{Tr} \left[ \mathbf B^P \mathbf T^{x} \right] \notag \\
&= 2 \text{Tr} \left[ \overline{\mathbf W}^P \mathbf S^x \right]
\end{align}
where $\overline{\mathbf W}^P$ is defined according to Eq. \ref{eq:W-bar}.

\subsubsection{Centroid distance matrix}
\begin{align} 
A_{x}^D &= \mathbf Q^D  \cdot \frac{\partial \mathbf D^\text{SAAO}}{\partial x}
\end{align}
with
\begin{align}
\frac{\partial D^\text{SAAO}_{\kappa \lambda}}{\partial x} = \frac{1}{D^\text{SAAO}_{\kappa \lambda}} \vec d_{\kappa \lambda} \cdot \frac{\partial \vec d_{\kappa \lambda}}{\partial x}
\end{align}
where $\frac{\partial \vec d_{\kappa \lambda}}{\partial x}$ is defined as
\begin{align}
\frac{\partial \vec d_{\kappa \lambda}}{\partial x} &= \left[ \left( \mathbf X_\kappa^\dagger \mathbf r^\text{AO} \frac{\partial \mathbf X_\kappa}{\partial x} - \mathbf X_\lambda^\dagger \mathbf r^\text{AO} \frac{\partial \mathbf X_\lambda}{\partial x} \right) + \text{transpose} \right] +  (\mathbf X^\dagger  \frac{\partial  \mathbf r^\text{AO}}{\partial x} \mathbf X)_{\kappa\kappa} - (\mathbf X^\dagger  \frac{\partial  \mathbf r^\text{AO}}{\partial x} \mathbf X)_{\lambda\lambda} \notag \\
&= 2 \left[ (\mathbf r^\text{SAAO} \mathbf T^{x})_{\kappa\kappa} - (\mathbf r^\text{SAAO} \mathbf T^{x})_{\lambda\lambda}\right] +  (\mathbf X^\dagger  \frac{\partial  \mathbf r^\text{AO}}{\partial x} \mathbf X)_{\kappa\kappa} - (\mathbf X^\dagger  \frac{\partial  \mathbf r^\text{AO}}{\partial x} \mathbf X)_{\lambda\lambda} 
\end{align}

This leads to
\begin{align} 
A_{x}^D &= \mathbf Q^D  \cdot \frac{\partial \mathbf D^\text{SAAO}}{\partial x} = \sum_{\kappa \lambda} Q^D_{\kappa \lambda} \frac{\partial D^\text{SAAO}_{\kappa \lambda}}{\partial x}  \notag \\
&=  \sum_{\kappa \lambda}  \frac{Q^D_{\kappa \lambda}}{D^\text{SAAO}_{\kappa \lambda}} \vec d_{\kappa \lambda} \cdot \frac{\partial \vec d_{\kappa \lambda}}{\partial x} \notag =  \sum_{\kappa \lambda} \vec R_{\kappa \lambda} \cdot \frac{\partial \vec d_{\kappa \lambda}}{\partial x} \notag \\
&= \sum_{\kappa \lambda} \vec R_{\kappa \lambda} \cdot \left( 2 \left[ (\mathbf r^\text{SAAO} \mathbf T^{x})_{\kappa\kappa} - (\mathbf r^\text{SAAO} \mathbf T^{x})_{\lambda\lambda}\right] +  (\mathbf X^\dagger  \frac{\partial  \mathbf r^\text{AO}}{\partial x} \mathbf X)_{\kappa\kappa} - (\mathbf X^\dagger  \frac{\partial  \mathbf r^\text{AO}}{\partial x} \mathbf X)_{\lambda\lambda} \right) \notag \\
&= 2 \sum_{\kappa \lambda} \vec R_{\kappa \lambda} \cdot \left( 2 (\mathbf r^\text{SAAO} \mathbf T^{x})_{\kappa\kappa} +  (\mathbf X^\dagger  \frac{\partial  \mathbf r^\text{AO}}{\partial x} \mathbf X)_{\kappa\kappa}  \right)
\end{align}
where $\vec R$ is defined in \cref{eq:R}.

Define $\bar{\mathbf d}^L$as 
\begin{align}
\bar{\mathbf d}^L_{\mu\nu} &= \sum_\kappa X_{\mu \kappa} X_{\nu \kappa} \vec R^L_{\kappa} 
\end{align}
where $\vec R^L$ is defined in \cref{eq:R-L}.

Then
\begin{align} 
A_{x}^D
&= 4 \text{Tr} [ \mathbf B^D \mathbf T^x ] + 2 \text{Tr} \left[ \bar{\mathbf d}^L \cdot  \frac{\partial  \mathbf r^\text{AO}}{\partial x} \right] \notag \\
&= 4 \text{Tr} \left[ \overline{\mathbf W}^D \mathbf S^x \right] + 2 \text{Tr} \left[ \bar{\mathbf d}^L \cdot  \frac{\partial  \mathbf r^\text{AO}}{\partial x} \right]. 
\end{align}

\subsection{xTB generalized Fock matrix}
The xTB generalized Fock matrix is defined as
\begin{align}
\left(\mathbf g[\mathbf Y] \right)_{\mu\nu} = \sum_{\kappa\lambda} \frac{\partial F_{\mu\nu}}{\partial P_{\kappa\lambda}}  Y_{\kappa\lambda}
\end{align}
where $\mathbf Y$ is an arbitrary symmetric matrix with the same dimension as the AO density matrix $\mathbf P$.

The xTB Fock matrix is defined as
\begin{align}
F_{\mu\nu} &= H_{\mu\nu} + \frac{1}{2} S_{\mu\nu} \sum_{C, l''} (\gamma_{AC, ll''} + \gamma_{BC, l'l''}) p^C_{l''} + \frac{1}{2} S_{\mu\nu} (q_A^2 \Gamma_A + q_B^2 \Gamma_B)  \quad (\mu \in A, l; \nu \in B, l') 
\end{align}
which is a functional of the shell-resolved charges, i.e. $\mathbf F[p^C_{l''}]$.

With the above expression, the xTB generalized Fock matrix can be computed as
\begin{align}
\left(\mathbf g[\mathbf Y] \right)_{\mu\nu} &= \sum_{\kappa\lambda} \frac{\partial F_{\mu\nu}}{\partial P_{\kappa\lambda}}  Y_{\kappa\lambda} =  \sum_{C, l''} \sum_{\kappa\lambda} \frac{\partial F_{\mu\nu}}{\partial p^C_{l''}}   \frac{\partial p^C_{l''}}{\partial P_{\kappa\lambda}} Y_{\kappa\lambda} 
\end{align}

The shell-resolved charges $p^C_{l''}$ are defined as
\begin{align}
p^C_{l''} = {p^C_{l''}}^0 - \sum_{\kappa \in C, l''} \sum_\lambda S_{\kappa\lambda} P_{\kappa\lambda} 
\end{align}
Define
\begin{align}
\tilde p^C_{l''} &\equiv \sum_{\kappa\lambda}  \frac{\partial p^C_{l''}}{\partial P_{\kappa\lambda}} Y_{\kappa\lambda} = - \sum_{\kappa \in C, l''} \sum_\lambda S_{\kappa\lambda} Y_{\kappa\lambda} 
\end{align}

The final expression for the xTB generalized Fock matrix is
\begin{align} \label{eq:gen_fock}
\left(\mathbf g[\mathbf Y] \right)_{\mu\nu} &= \sum_{C, l''} \sum_{\kappa\lambda} \frac{\partial F_{\mu\nu}}{\partial p^C_{l''}}   \frac{\partial p^C_{l''}}{\partial P_{\kappa\lambda}} Y_{\kappa\lambda} \notag = \sum_{C, l''}  \frac{\partial F_{\mu\nu}}{\partial p^C_{l''}}  \tilde p^C_{l''} \notag \\
&=  \frac{1}{2} S_{\mu\nu} \sum_{C, l''} (\gamma_{AC, ll''} + \gamma_{BC, l'l''}) \tilde p^C_{l''} + S_{\mu\nu} (q_A \tilde q_A \Gamma_A + q_B \tilde q_B \Gamma_B) 
\end{align}
where $\tilde q_A = \sum_{l} \tilde p^A_{l}$.

\subsection{Coupled-perturbed z-vector equation for xTB}
Combining the stationary condition of the Lagrangian, Eq. \ref{eq:stationary_cond} and the condition $\mathbf x = \mathbf x^\dagger$ leads to the coupled-perturbed z-vector equation for xTB:
\begin{align}
(\varepsilon_a - \varepsilon_i) z_{ai} + 2[\mathbf g(\bar{\mathbf z})]_{ai} = -(A_{ai} - A_{ia})
\end{align}
where $\varepsilon_a, \varepsilon_i$ are the xTB orbital energies, $\mathbf z$ is the Lagrange multiplier defined in Eq. \ref{eq:L}. $\bar{\mathbf z} = \mathbf z + \mathbf z^\dagger$.

$\mathbf g(\bar{\mathbf z})$ is the generalized xTB Fock matrix and is defined in Eq. \ref{eq:gen_fock}.

\subsection{Expression for $\mathbf W$}
The stationary condition of the Lagrangian, Eq. \ref{eq:stationary_cond} also leads to the expression for the weight matrix
$\mathbf W$:
\begin{align}
W_{pq} = -\frac{1}{4} (1 + \hat P_{pq}) [\mathbf A + \mathbf A(\mathbf z)]_{pq}
\end{align}
where $\hat P_{pq}$ is the permutation operator that permutes indices $p$ and $q$.

\subsection{Final gradient expression}
With all intermediate quantities obtained in the previous sections, we can now write the expression for the OrbNet energy gradient:
\begin{align}
\frac{d E_{\text{out}}}{d x} = \frac{\partial E_{\text{out}}}{\partial x}  + \text{Tr} [\mathbf W \mathbf{S}^x]   +  \text{Tr} [\mathbf  z \mathbf F^{(x)}]  \label{eq:L_grad}
\end{align}
where the first term on the right-hand-side can be computed as
\begin{align}
\frac{\partial E_{\text{out}}}{\partial x} &= \frac{d E_{\text{xTB}}}{d x} + \sum_f \left[ \mathbf Q^f \frac{\partial \mathbf f}{\partial x} \right]
\end{align}

\begin{align}
\frac{d E_{\text{out}}}{d x} &= \frac{d E_{\text{xTB}}}{d x}  + \sum_f \left[ \mathbf Q^f \frac{\partial \mathbf f}{\partial x} \right] + \text{Tr} [\mathbf W \mathbf{S}^x]   +  \text{Tr} [\mathbf  z \mathbf F^{(x)}]  \notag \\
&= \frac{d E_{\text{xTB}}}{d x}  + \sum_f \left[ \mathbf Q^f \frac{\partial \mathbf f}{\partial x} \right] + \text{Tr} [\mathbf W \mathbf{S}^x]   +  \text{Tr} [\mathbf  z^\text{AO} \frac{\partial \mathbf F^{\text{AO}}}{\partial x}] \notag \\
&= \frac{d E_{\text{xTB}}}{d x} + \text{Tr} [\mathbf W \mathbf{S}^x]   +  \text{Tr} [\mathbf  z^\text{AO} \mathbf F^x] \notag \\
&\quad + 2 \text{Tr} \left[ \overline{\mathbf W}^H \mathbf S^x \right] + \text{Tr} \left[ \mathbf Q^{H,\text{AO}} \mathbf H^x \right] \notag \\
&\quad + 2 \text{Tr} \left[ \overline{\mathbf W}^S \mathbf S^x \right] + \text{Tr} \left[ \mathbf Q^{S, \text{AO}} \mathbf S^x  \right] \notag \\
&\quad + 2 \text{Tr} \left[ \overline{\mathbf W}^F \mathbf S^x \right]  + \text{Tr} \left[ \mathbf Q^{F, \text{AO}} \mathbf F^x \right] \notag \\
&\quad + 2 \text{Tr} \left[ \overline{\mathbf W}^P \mathbf S^x \right] \notag 
\quad + 4 \text{Tr} \left[ \overline{\mathbf W}^D \mathbf S^x \right] + 2\text{Tr} \left[\bar{\mathbf d}^L \cdot  \mathbf r^x \right] \notag \\
\label{eq:L_grad}
\end{align}

The GFN-xTB gradient is written as \cite{gfn1}
\begin{align}
\frac{d E_{\text{xTB}}}{d x} &= \text{Tr} [ \mathbf P \mathbf H^x ] + E_\text{h2}^x + E_\text{h3}^x
\end{align}

\section{Auxiliary basis set for density matrix projection}
\label{sec:aux_proj}

The basis set file used to produced the projected density matrix auxiliary targets, reported in the NWChem format:

\scriptsize
\#BASIS SET: (30,30p,30d) -> [30,30p,30d]\\
HCONFSCl SPD\\
    2.560000000000e+02 1.0 -1.0 0.0 0.0 0.0 0.0 0.0 0.0 0.0 0.0 0.0 0.0 0.0 0.0 0.0 0.0 0.0 0.0 0.0 0.0 0.0 0.0 0.0 0.0 0.0 0.0 0.0 0.0 0.0 0.0\\
    1.280000000000e+02 0.0 1.0 -1.0 0.0 0.0 0.0 0.0 0.0 0.0 0.0 0.0 0.0 0.0 0.0 0.0 0.0 0.0 0.0 0.0 0.0 0.0 0.0 0.0 0.0 0.0 0.0 0.0 0.0 0.0 0.0\\
    6.400000000000e+01 0.0 0.0 1.0 -1.0 0.0 0.0 0.0 0.0 0.0 0.0 0.0 0.0 0.0 0.0 0.0 0.0 0.0 0.0 0.0 0.0 0.0 0.0 0.0 0.0 0.0 0.0 0.0 0.0 0.0 0.0\\
    3.200000000000e+01 0.0 0.0 0.0 1.0 -1.0 0.0 0.0 0.0 0.0 0.0 0.0 0.0 0.0 0.0 0.0 0.0 0.0 0.0 0.0 0.0 0.0 0.0 0.0 0.0 0.0 0.0 0.0 0.0 0.0 0.0\\
    1.600000000000e+01 0.0 0.0 0.0 0.0 1.0 -1.0 0.0 0.0 0.0 0.0 0.0 0.0 0.0 0.0 0.0 0.0 0.0 0.0 0.0 0.0 0.0 0.0 0.0 0.0 0.0 0.0 0.0 0.0 0.0 0.0\\
    8.000000000000e+00 0.0 0.0 0.0 0.0 0.0 1.0 -1.0 0.0 0.0 0.0 0.0 0.0 0.0 0.0 0.0 0.0 0.0 0.0 0.0 0.0 0.0 0.0 0.0 0.0 0.0 0.0 0.0 0.0 0.0 0.0\\
    4.000000000000e+00 0.0 0.0 0.0 0.0 0.0 0.0 1.0 -1.0 0.0 0.0 0.0 0.0 0.0 0.0 0.0 0.0 0.0 0.0 0.0 0.0 0.0 0.0 0.0 0.0 0.0 0.0 0.0 0.0 0.0 0.0\\
    2.000000000000e+00 0.0 0.0 0.0 0.0 0.0 0.0 0.0 1.0 -1.0 0.0 0.0 0.0 0.0 0.0 0.0 0.0 0.0 0.0 0.0 0.0 0.0 0.0 0.0 0.0 0.0 0.0 0.0 0.0 0.0 0.0\\
    1.333333333333e+00 0.0 0.0 0.0 0.0 0.0 0.0 0.0 0.0 1.0 -1.0 0.0 0.0 0.0 0.0 0.0 0.0 0.0 0.0 0.0 0.0 0.0 0.0 0.0 0.0 0.0 0.0 0.0 0.0 0.0 0.0\\
    1.000000000000e+00 0.0 0.0 0.0 0.0 0.0 0.0 0.0 0.0 0.0 1.0 -1.0 0.0 0.0 0.0 0.0 0.0 0.0 0.0 0.0 0.0 0.0 0.0 0.0 0.0 0.0 0.0 0.0 0.0 0.0 0.0\\
    8.000000000000e-01 0.0 0.0 0.0 0.0 0.0 0.0 0.0 0.0 0.0 0.0 1.0 -1.0 0.0 0.0 0.0 0.0 0.0 0.0 0.0 0.0 0.0 0.0 0.0 0.0 0.0 0.0 0.0 0.0 0.0 0.0\\
    6.666666666667e-01 0.0 0.0 0.0 0.0 0.0 0.0 0.0 0.0 0.0 0.0 0.0 1.0 -1.0 0.0 0.0 0.0 0.0 0.0 0.0 0.0 0.0 0.0 0.0 0.0 0.0 0.0 0.0 0.0 0.0 0.0\\
    5.714285714286e-01 0.0 0.0 0.0 0.0 0.0 0.0 0.0 0.0 0.0 0.0 0.0 0.0 1.0 -1.0 0.0 0.0 0.0 0.0 0.0 0.0 0.0 0.0 0.0 0.0 0.0 0.0 0.0 0.0 0.0 0.0\\
    5.000000000000e-01 0.0 0.0 0.0 0.0 0.0 0.0 0.0 0.0 0.0 0.0 0.0 0.0 0.0 1.0 -1.0 0.0 0.0 0.0 0.0 0.0 0.0 0.0 0.0 0.0 0.0 0.0 0.0 0.0 0.0 0.0\\
    4.444444444444e-01 0.0 0.0 0.0 0.0 0.0 0.0 0.0 0.0 0.0 0.0 0.0 0.0 0.0 0.0 1.0 -1.0 0.0 0.0 0.0 0.0 0.0 0.0 0.0 0.0 0.0 0.0 0.0 0.0 0.0 0.0\\
    4.000000000000e-01 0.0 0.0 0.0 0.0 0.0 0.0 0.0 0.0 0.0 0.0 0.0 0.0 0.0 0.0 0.0 1.0 -1.0 0.0 0.0 0.0 0.0 0.0 0.0 0.0 0.0 0.0 0.0 0.0 0.0 0.0\\
    3.636363636364e-01 0.0 0.0 0.0 0.0 0.0 0.0 0.0 0.0 0.0 0.0 0.0 0.0 0.0 0.0 0.0 0.0 1.0 -1.0 0.0 0.0 0.0 0.0 0.0 0.0 0.0 0.0 0.0 0.0 0.0 0.0\\
    3.333333333333e-01 0.0 0.0 0.0 0.0 0.0 0.0 0.0 0.0 0.0 0.0 0.0 0.0 0.0 0.0 0.0 0.0 0.0 1.0 -1.0 0.0 0.0 0.0 0.0 0.0 0.0 0.0 0.0 0.0 0.0 0.0\\
    3.076923076923e-01 0.0 0.0 0.0 0.0 0.0 0.0 0.0 0.0 0.0 0.0 0.0 0.0 0.0 0.0 0.0 0.0 0.0 0.0 1.0 -1.0 0.0 0.0 0.0 0.0 0.0 0.0 0.0 0.0 0.0 0.0\\
    2.857142857143e-01 0.0 0.0 0.0 0.0 0.0 0.0 0.0 0.0 0.0 0.0 0.0 0.0 0.0 0.0 0.0 0.0 0.0 0.0 0.0 1.0 -1.0 0.0 0.0 0.0 0.0 0.0 0.0 0.0 0.0 0.0\\
    2.666666666667e-01 0.0 0.0 0.0 0.0 0.0 0.0 0.0 0.0 0.0 0.0 0.0 0.0 0.0 0.0 0.0 0.0 0.0 0.0 0.0 0.0 1.0 -1.0 0.0 0.0 0.0 0.0 0.0 0.0 0.0 0.0\\
    2.500000000000e-01 0.0 0.0 0.0 0.0 0.0 0.0 0.0 0.0 0.0 0.0 0.0 0.0 0.0 0.0 0.0 0.0 0.0 0.0 0.0 0.0 0.0 1.0 -1.0 0.0 0.0 0.0 0.0 0.0 0.0 0.0\\
    2.352941176471e-01 0.0 0.0 0.0 0.0 0.0 0.0 0.0 0.0 0.0 0.0 0.0 0.0 0.0 0.0 0.0 0.0 0.0 0.0 0.0 0.0 0.0 0.0 1.0 -1.0 0.0 0.0 0.0 0.0 0.0 0.0\\
    2.222222222222e-01 0.0 0.0 0.0 0.0 0.0 0.0 0.0 0.0 0.0 0.0 0.0 0.0 0.0 0.0 0.0 0.0 0.0 0.0 0.0 0.0 0.0 0.0 0.0 1.0 -1.0 0.0 0.0 0.0 0.0 0.0\\
    2.105263157895e-01 0.0 0.0 0.0 0.0 0.0 0.0 0.0 0.0 0.0 0.0 0.0 0.0 0.0 0.0 0.0 0.0 0.0 0.0 0.0 0.0 0.0 0.0 0.0 0.0 1.0 -1.0 0.0 0.0 0.0 0.0\\
    2.000000000000e-01 0.0 0.0 0.0 0.0 0.0 0.0 0.0 0.0 0.0 0.0 0.0 0.0 0.0 0.0 0.0 0.0 0.0 0.0 0.0 0.0 0.0 0.0 0.0 0.0 0.0 1.0 -1.0 0.0 0.0 0.0\\
    1.904761904762e-01 0.0 0.0 0.0 0.0 0.0 0.0 0.0 0.0 0.0 0.0 0.0 0.0 0.0 0.0 0.0 0.0 0.0 0.0 0.0 0.0 0.0 0.0 0.0 0.0 0.0 0.0 1.0 -1.0 0.0 0.0\\
    1.818181818182e-01 0.0 0.0 0.0 0.0 0.0 0.0 0.0 0.0 0.0 0.0 0.0 0.0 0.0 0.0 0.0 0.0 0.0 0.0 0.0 0.0 0.0 0.0 0.0 0.0 0.0 0.0 0.0 1.0 -1.0 0.0\\
    1.739130434783e-01 0.0 0.0 0.0 0.0 0.0 0.0 0.0 0.0 0.0 0.0 0.0 0.0 0.0 0.0 0.0 0.0 0.0 0.0 0.0 0.0 0.0 0.0 0.0 0.0 0.0 0.0 0.0 0.0 1.0 -1.0\\
    1.666666666667e-01 0.0 0.0 0.0 0.0 0.0 0.0 0.0 0.0 0.0 0.0 0.0 0.0 0.0 0.0 0.0 0.0 0.0 0.0 0.0 0.0 0.0 0.0 0.0 0.0 0.0 0.0 0.0 0.0 0.0 1.0
\normalsize

\printbibliography[title={Appendix References}]

\end{refsection}

\end{document}